\documentclass[useAMS, usenatbib, usegraphicx]{mn2e} 
\usepackage{rotating} 

\title[Kinetic AGN Feedback] 
{Kinetic or thermal AGN feedback in simulations of isolated and merging disk galaxies 
calibrated by the $M - \sigma$ relation} 

\author[P. Barai et al.] 
{Paramita Barai$^{1}$\thanks{E-mail: pbarai@oats.inaf.it}, 
Matteo Viel$^{1, 2}$, 
Giuseppe Murante$^{1}$, 
Massimo Gaspari$^{3}$, 
\newauthor 
Stefano Borgani$^{4, 1, 2}$ 
\vspace{0.2cm} \\ 
$^{1}$ INAF - Osservatorio Astronomico di Trieste, Via G.B. Tiepolo 11, I-34143 Trieste, Italy \\ 
$^{2}$ INFN / National Institute for Nuclear Physics, Via Valerio 2, I-34127 Trieste, Italy \\ 
$^{3}$ Max Planck Institute for Astrophysics, Karl-Schwarzschild-Strasse 1, D-85741 Garching, Germany \\ 
$^{4}$ Dipartimento di Fisica dell'Universit\`{a} di Trieste, Sezione di Astronomia, 
Via Tiepolo 11, I-34131 Trieste, Italy \\ 
} 

\begin{document} 

\maketitle

\label{firstpage} 

\begin{abstract} 

We investigate two modes of coupling the feedback energy from a central AGN 
to the neighbouring gas in galaxy simulations: 
{\it kinetic} - velocity boost, and {\it thermal} - heating. 
We formulate kinetic feedback models for energy-driven wind (EDW) and momentum-driven wind (MDW), 
using two free parameters: feedback efficiency $\epsilon_f$, and AGN wind velocity $v_w$. 
A novel numerical algorithm is implemented in the SPH code {\sc GADGET-3}, 
to prevent the expansion of a hole in the gas distribution around the BH. 
We perform simulations of isolated evolution and merger of disk galaxies, 
of Milky-Way mass as well as lower and higher masses. 
We find that in the isolated galaxy BH kinetic feedback generates intermittent bipolar jet-like gas outflows. 
We infer that current prescriptions for BH subgrid physics in galaxy simulations 
can grow the BH to observed values even in an isolated disk galaxy. 
The BH growth is enhanced in a galaxy merger, 
which consequently requires different model parameters to fit the observations than an isolated case. 
Comparing the $[M_{\rm BH} - \sigma_{\star}]$ relation 
obtained in our simulations with observational data, 
we conclude that it is possible to find parameter sets for a fit in all the models 
(e.g. $v_w = 10000$ km/s and $\epsilon_f = 0.25$ for BH kinetic EDW), 
except for the case with MDW feedback in a galaxy merger, in which the BH is always too massive. 
The BH thermal feedback implementation of \citet{SDH05} 
within the multiphase star-formation model is found to have negligible impact on gas properties; 
and the effect claimed in all previous studies is attributed to 
gas depletion around the BH by the creation of an artificial hole. 
The BH mass accretion rate in our simulations exhibit heavy fluctuations. 
The star formation rate is quenched with feedback by removal of gas. 
The circumgalactic medium (CGM) gas at galactocentric distances $(20 - 100) h^{-1}$ kpc 
are found to give the best metallicity observational diagnostic to distinguish between BH models. 

% Therefore full convergence cannot be expected with the same parameter values of the subgrid models. 
% Rigorous numerical convergence would first require tuning the SF threshold density and SF efficiency. 

% visible shortly. 
% of amplitude by up to factors $100 - 1000$. 
% In the galaxy merger, the star formation rate (SFR) is suppressed 
% by $(1.5 - 10)$ times with thermal feedback, and by $(5 - 100)$ times with kinetic feedback. 
% that we limit using our novel numerical technique, 
% and it is difficult to perform a physical comparison with kinetic feedback. 
% and $0.03 - 0.07$ of the total gas mass of galaxy disk outflows by $1$ Gyr. 
% and there is very little effect of feedback by gas heating on SFR.  
% We include metal-dependent radiative cooling and heating, star formation, stellar evolution, 
% chemical enrichment, and vary the AGN feedback models. 
% where kinetic feedback results up to $(10 - 1000)$ times higher $Z_C$ than thermal. 
% of \citet{SH03}, 
% where SF is based on a density threshold only. 
% where the gas is % Our results imply that 
% suitable % combinations % is significantly different between the models 
% The BH mass undergoes 
% an exponential growth from $0.5$ Gyr until $(1 - 1.5)$ Gyr, and becomes almost steady-state afterward. 

\end{abstract} 


\section{Introduction} 
\label{sec-intro} 

Supermassive black holes (SMBHs) are believed to exist at the 
centers of active galaxies \citep[e.g.,][]{Kormendy95, Ferrarese05}, 
powered by the accretion of matter and liberating enormous amounts of energy. 
Active Galactic Nuclei (AGN) influence the formation and evolution of galaxies in the form of feedback 
\citep[e.g.,][]{Silk98, King03, Granato04, Begelman05, Barai06, Croton06, Barai08, Fabian12, Wagner13}, 
generating observational trends such as the central SMBH - host galaxy stellar bulge correlations 
\citep[e.g.,][]{Magorrian98, Gebhardt00, Shankar06}. 
The energy output is often observed as AGN outflows in a wide variety of forms 
\citep[see][for reviews]{Crenshaw03, Everett07}, e.g.: 
collimated relativistic jets and/or huge overpressured cocoons in radio \citep{Nesvadba08}, 
blue-shifted broad absorption lines in the UV and optical \citep{Reichard03, Rupke11}, 
warm absorbers \citep{Krongold07} and ultra-fast outflows \citep{Tombesi13} in X-rays, 
molecular gas in far-IR \citep{Feruglio10, Sturm11}. 

Concordance galaxy formation models in the cold dark matter cosmology 
widely incorporate feedback from AGN in simulations of 
isolated galaxies and mergers \citep[e.g.,][]{SDH05, Johansson09b}, 
and cosmological volumes 
\citep[e.g.,][]{Sijacki07, Booth09, Dubois10, Fabjan10, Barai11a, DiMatteo12, Hirschmann13}; 
as well as semi-analytical studies \citep[e.g.,][]{Salucci99, Shankar04}. 
Simulations generally invoke AGN feedback in the negative form which quenches star formation 
and limits the formation of massive stellar systems \citep[e.g.,][]{Scannapieco05, vandeVoort11, Dubois13}, 
as supported by some observations \citep[e.g.,][]{Schawinski06, Wang07}. 
At the same time, AGN feedback can be positive occasionally which also plays an important role 
in the cosmological context. 
AGN outflows has been shown to overpressure, compress and fragment clumpy gas clouds, triggering starbursts, 
in theoretical and numerical studies \citep[e.g.,][]{DeYoung89, Silk05, Zubovas13}, 
and observed in jet-induced star formation and radio-optical alignment \citep[e.g.,][]{Chambers87, Zinn13}. 

% enhanced star formation rate of radio selected AGN arises because of 

We investigate, in this paper, different models and implementation of AGN feedback in galaxy simulations. 
Our goal is to compare and contrast 
two modes of coupling of the feedback energy from BH to the surrounding gas: 
{\it thermal} - where the gas temperature (or, internal energy) is increased, and 
{\it kinetic} - where the gas velocity is boosted. 
We aim to find the model parameters in each case which fits relevant observational data, 
and explore the signatures of various feedback models on the BH growth, galaxy and CGM properties. 
In this work we perform simulations of isolated and merging disk galaxies, 
before applying our models to cosmological volumes in the future. 

Galaxy formation simulations have investigated  both AGN feedback mechanisms: 
thermal \citep[e.g.,][]{DiMatteo05, Sijacki07, Booth09, Fabjan10, Gaspari11b}, 
and kinetic \citep[e.g.,][]{Germain09, Dubois10, Ostriker10, Gaspari12a, Vazza13}. 
\citet{Gaspari11a} explored several feedback mechanisms in galaxy clusters, including cold gas accretion 
and massive subrelativistic outflows, that self-regulate the mechanical power from AGN outflow heating. 
These models were extended by \citet{Gaspari11b} to galaxy groups. 
\citet{Gaspari12a} studied the role of mechanical AGN feedback in 
controlling the thermodynamical evolution of isolated elliptical galaxies. 
\citet{Gaspari12b} explored the formation of multiphase gas via thermal instability 
in cluster cores heated by collimated bipolar AGN jets. 

There has been two contemporary studies of BH feedback in isolated galaxy systems, 
the distinction of our work from those is outlined below. 
\citet{Newton13} simulated isolated and merging disc galaxies 
to investigate the effect of feedback from both AGN and supernovae on galaxy evolution, 
and to isolate the most important factors of these feedback processes. 
They utilize different methods for distributing the feedback energy in the same thermal form, 
and do not have kinetic feedback. 
Our work considers the kinetic mode in addition to thermal. 

\citet{Wurster13b} compared the AGN feedback algorithms of four authors 
\citep{SDH05, Okamoto08, Booth09, Debuhr11} together with their own model in galaxy merger simulations, 
and found wide variation in accretion behaviours. 
Among these, in \citet{Debuhr11} the feedback is returned as momentum 
or in the kinetic form, while the others have thermal feedback. 
However the model by \citet{Debuhr11} also has a distinct accretion prescription 
based on the viscous transport of angular momentum; 
therefore comparing it with e.g. \citet{SDH05} (which uses the modified Bondi accretion rate) 
makes it hard to disentangle if the differential effects 
are because of various accretion methods or varying feedback models. 
In our study we use the same accretion methodology 
and then compare between thermal versus kinetic modes of feedback. 

This paper is organised as follows: 
we describe our numerical code and simulation setup in \S\ref{sec-numerical}, 
in \S\ref{sec-results} we present and analyse our results, 
some important outcomes are discussed in \S\ref{sec-discussion}, 
while in \S\ref{sec-conclusion} we give a summary of the main findings 
and discuss possible future applications.

\section{Numerical Method} 
\label{sec-numerical} 

We use a modified version of the TreePM (particle mesh) - 
SPH (smoothed particle hydrodynamics) code {\sc GADGET-3} \citep{Springel05}, 
which includes a more efficient domain decomposition to improve the work-load balance 
over the public version {\sc GADGET-2}. 
Some of the subgrid\footnote{By {\it subgrid} we mean 
{\it sub-resolution}, referring to physical processes occurring at length scales 
smaller than the resolved scales in our simulations.} 
physics included in the 
semi-public version of {\sc GADGET-3} code we use are outlined below. 
The BH modules including our new kinetic feedback model are described in 
\S\ref{sec-num-BH-Accr-Feed}, \S\ref{sec-num-BH-Kin} and \S\ref{sec-num-Implement}. 
The initial galaxy models are presented in \S\ref{sec-num-IC}, and 
our simulations are mentioned in \S\ref{sec-num-Sim}. 

The non-AGN subgrid models: radiative physics, star-formation (SF) and chemical evolution, 
are same as the work of \citet{Barai13}. 
Radiative cooling and heating is computed by adding metal-line cooling from \citet{Wiersma09a}, 
considering 11 different elements: H, He, C, Ca, O, N, Ne, Mg, S, Si, Fe. 
A photoionizing background radiation from the cosmic microwave background (CMB) 
and the \citet{Haardt01} model for the UV/X-ray background are considered. 

SF is implemented following the multiphase effective subresolution model by \citet{SH03}. 
Gas particles with density above a limiting threshold, $\rho_{\rm SF} = 0.13$ cm$^{-3}$ 
(units of number density of hydrogen atoms), 
contain cold and hot phases, and are star forming. 
Collisionless star particles are spawned from gas particles undergoing SF, 
based on the stochastic scheme by \citet{Katz96}. 
We allow a gas particle to spawn up to four generations of stars. 

Stellar evolution and chemical enrichment feedback 
are incorporated following the chemical evolution model of \citet{Tornatore07}. 
Production of 11 species (H, He, C, Ca, O, N, Ne, Mg, S, Si, Fe) 
are accounted for using detailed yields from Type Ia SN (SN-Ia), Type II SN (SN-II), 
along with low and intermediate mass stars (LIMS) 
in the thermally pulsating asymptotic giant branch (TP-AGB) phase. 
Both SN-Ia and SN-II contributes to thermal feedback. 
There are mass-dependent time delays with which different stellar populations release metals, 
adopting the lifetime function by \citet{Padovani93}. 
Different stellar yields are used: for SN-Ia taken from \citet{Thielemann03}, 
SN-II from \citet{Woosley95}, and LIMS from \citet{vandenHoek97}. 
The mass range for SN-II is considered to be $M / M_{\odot} > 8$, 
while that for SN-Ia originating from binary systems is $0.8 < M / M_{\odot} < 8$ 
with a binary fraction of $10\%$. 

We include a fixed stellar initial mass function (IMF) according to the formalism given by 
\citet{Chabrier03}, which is a power-law at $M / M_{\odot} > 1$ and has a log-normal form at masses below. 
However, we use power-law IMFs with different slopes 
over the whole mass range of $0.1$ to $100 M_{\odot}$, 
which mimics the log-normal form of \citet{Chabrier03} at lower masses, as tests indicate. 
The functional form: $\phi \left( M \right) = K M^{-y}$, 
is composed of 3 slopes and normalizations: 
$y = 0.2$ and $K = 0.497$ for stellar masses $0.1 \leq M / M_{\odot} < 0.3$, 
$y = 0.8$ and $K = 0.241$ for $0.3 \leq M / M_{\odot} < 1$, and 
$y = 1.3$ and $K = 0.241$ for $1 \leq M / M_{\odot} < 100$. 
Stars within a mass interval $[8 - 40] M_{\odot}$ 
become SNe first before turning into black holes at the end of their lives, 
while stars of mass $> 40 M_{\odot}$ are allowed to directly end in BHs. 

The chemical evolution model incorporates mass loss through stellar winds and SNe explosions, 
which are self-consistently computed for a given IMF and lifetime function. 
A fraction of a star particle's mass is restored as diffuse gas 
during its evolution, and distributed to the surrounding gas particles. 
There is no kinetic feedback from SNe-driven galactic outflows in our simulations. 
This is because we want to decouple AGN-driven from SNe-driven outflowing gas, 
and aim to explore uncontaminated outflows driven solely by thermal or kinetic AGN feedback. 

% The code contains 
% a time integration using energy and entropy conserving formulation of SPH \citep{Springel02}, 
% uses fully adaptive smoothing lengths, 
% and a standard SPH artificial viscosity prescription \citep{Monaghan97}. 
% {\sc GADGET-3} additionally % numerical models 
% based on the original prescription of \citet{Katz96}, and improved 
% Net cooling rates are computed element-by-element tracking 11 species: 
% spatially-uniform time-dependent 
% therefore a typical star particle mass is about one-fourth of the initial mass of gas particles. 

% Both SN-Ia and SN-II contribute to thermal feedback, 
% which essentially has no effect in the effective multiphase star-formation model, 
% except an impact of SN-Ia heating up relatively lower density gas. 

\subsection{BH Accretion and Energy Feedback} 
\label{sec-num-BH-Accr-Feed} 

Our subgrid models for BH accretion and feedback are based on the 
original prescriptions by \citet{SDH05}, which we extend to include kinetic feedback. 
The mass inflow rate of surrounding gas onto a central SMBH of mass $M_{\rm BH}$ is parametrized 
by the rate given by \citet{Hoyle39, Bondi44, Bondi52}: 
\begin{equation} 
\label{eq-Mdot-Bondi} 
\dot{M}_{\rm Bondi} = \alpha \frac{4 \pi G^2 M_{\rm BH}^2 \rho_{\infty}}{ \left(c_{s,\infty}^2 + v^2\right) ^ {3/2}}. 
\end{equation} 
In the original Bondi-Hoyle-Lyttleton parametrization, 
$\rho_{\infty}$ is the gas density far from the BH (or, at infinity), 
$c_{s,\infty}$ is the sound speed in the gas far from BH, 
$v$ is the velocity of the BH relative to the far-off gas, 
and the parameter $\alpha$ is analytically dependent on gas adiabatic index ($\gamma$) 
with $\alpha = 1 / 4$ for $\gamma = 5/3$. 
It was originally used to formulate critical astrophysical accretion, 
where gas is subsonic far away, passes through a sonic point, 
and accretes onto the central object with a supersonic velocity. 

% and high temperatures % multiphase structure of the ISM could be 

The issue of computational resolution appears: 
current standard galaxy formation simulations resolve kpc to 100's of pc length scales, 
hence the Bondi radius and sonic point ($\sim 10$'s of pc) are unresolved. 
The gas properties ($\rho_{\infty}, c_{s,\infty}$) used in Eq.~(\ref{eq-Mdot-Bondi}) 
are estimated by smoothing on the resolution scale (smoothing length $>$ a few $100$ pc) at the BH location. 
This results in artificially low densities compared to spatially resolving the Bondi radius scale. 
Furthermore, smaller-scale simulations \citep{Barai12, Gaspari13} 
show that the cooling gas is multiphase, with a variable accretion rate. 
This cold phase of the ISM is not resolved in galaxy simulations. 
As a numerical correction, the Bondi-Hoyle-Lyttleton accretion rate 
inferred from simulations is enhanced by setting the multiplicative factor to $\alpha \sim 100$ 
\citep[e.g.,][]{SDH05, Sijacki07, DiMatteo08, Khalatyan08, Johansson09a, Sijacki09, Dubois10}. 
\citet{Booth09} used an $\alpha$-factor dependent on the gas density. 

The idealized assumptions of the original Bondi theory: 
spherically-symmetric, non-rotating, adiabatic, steady and unperturbed gas flow 
with constant boundary conditions, 
has led to recent criticisms of the adopted Bondi accretion model in galaxy simulations. 
Using analytical arguments and simulating spherical gas distribution 
within the length scales $(0.001 - 1)$ kpc, \citet{Hobbs12} showed that 
in free-falling gas due to efficient cooling and gravity of the surrounding halo, 
the Bondi-Hoyle formalism can be erroneous by orders of magnitude in either direction. 
In sub-pc resolution simulations where the gas is cooling, 
\citet{Gaspari13} and \citet{Barai12} saw the formation of a multiphase medium, 
composed of thermal-instability driven cold clouds and filaments within the hot gas, 
which makes the accretion cold and chaotic. 
\citet{Gaspari13} inferred that the accretion rate 
is boosted up to two orders of magnitude compared with the Bondi prediction. 
Such ongoing studies are attempting to improve the BH accretion prescription 
in intermediate-scale simulations resolving the sonic radius \citep[also][]{Barai11b}. 
Alternate methods have also been prescribed recently to estimate the BH accretion rate on galaxy scales: 
use viscous transport of angular momentum \citep{Debuhr10}, 
accretion disc particle method \citep{Power11, Wurster13a}. 
Incorporating such modified accretion schemes in full cosmological simulations 
make up avenues for future work. 

% , due to the simplicity of its numerical implementation. 

Nevertheless despite the limitations, the Bondi model for BH accretion is widely used 
in galaxy-scale numerical studies, as we do in this work. 
The accretion rate estimated from the simulations must allow the 
BHs to grow from seed masses to that observable in the Universe today, 
within a few Gyrs or the Hubble time. 
It can be analytically shown that a BH, embedded in star-forming gas (having $\rho > \rho_{\rm SF}$) 
governed by the \citet{SH03} effective equation-of-state (\S\ref{sec-numerical}), 
accreting via Eq.~(\ref{eq-Mdot-Bondi}) has a mass growth time $t_g \propto 1 / \alpha$. 
The time $t_g$ is less than the Hubble time when $\alpha \geq 100$. 
With a smaller $\alpha$ value, $t_g$ exceeds the Hubble time and then 
it is never possible to grow the BH to the observed masses. 
Following these arguments to mimic the appropriate growth of BHs in our simulations, 
we adopt the Bondi-Hoyle-Lyttleton formulation with a constant multiplicative factor $\alpha = 100$. 

Furthermore accretion is limited to the Eddington rate, making the resultant BH mass accretion rate, 
\begin{equation} 
\label{eq-Mdot-BH} 
\dot{M}_{\rm BH} = {\rm min} \left( \dot{M}_{\rm Bondi}, \dot{M}_{\rm Edd} \right). 
\end{equation} 
Here, $\dot{M}_{Edd}$ is the Eddington mass accretion rate, expressed in terms of the Eddington luminosity, 
\begin{equation} 
\label{eq-LEdd} 
L_{\rm Edd} = \frac{4 \pi G M_{\rm BH} m_p c} {\sigma_T} = \epsilon_r \dot{M}_{\rm Edd} c^2. 
\end{equation} 

A fraction of the accreted rest-mass energy is considered to be radiated away by the  BH, 
assuming radiatively efficient accretion. The radiation luminosity is, 
\begin{equation} 
\label{eq-Lr-BH} 
L_r = \epsilon_r \dot{M}_{\rm BH} c^2, 
\end{equation} 
with $\epsilon_r$ being the radiative efficiency fraction. 
We adopt the mean value for radiatively efficient accretion onto a Schwarzschild BH 
\citep{Shakura73}: $\epsilon_r = 0.1$, which is kept fixed. 
This is supported by recent sub-pc resolution simulations of \citet{Maio13}, 
who found that possible values of radiative efficiencies should be between $0.09 - 0.15$. 

A fraction $\epsilon_f$ of the radiated energy is eventually fed back 
to the neighbouring gas as feedback energy from the BH: 
\begin{equation} 
\label{eq-Edot-Feed} 
\dot{E}_{\rm feed} = \epsilon_f L_r = \epsilon_f \epsilon_r \dot{M}_{\rm BH} c^2. 
\end{equation} 
Using $\epsilon_f = 0.05$, 
\citet{DiMatteo05} found consistent correlation of BH mass and 
galaxy stellar velocity dispersion (the $M_{\rm BH} - \sigma_{\star}$ relation), 
between galaxy merger simulations and observations. 
We consider the feedback efficiency $\epsilon_f$ as a free parameter in our models. 

We examine two ways in which the BH feedback energy is coupled to the surrounding gas: 

\noindent $\bullet$ {\bf Thermal} : 
We adopt the default scheme from \citet{SDH05}. 
The energy $\dot{E}_{\rm feed}$ is distributed thermally to heat up the gas isotropically around the BH. 
The temperature of the neighbouring gas particles 
(those contributing to Eq.~\ref{eq-BH-Smooth} in \S\ref{sec-num-Implement}) 
are incremented by an amount scaled by their SPH kernel weights. 
For a gas particle dense enough to be multiphase star-forming, 
the excess specific thermal energy decays to attain the specific energy of the 
effective equation-of-state, on a relaxation timescale $\tau_h$ 
(from Eq.~12 of \citealt{SH03}): 
\begin{equation} 
\tau_h = \frac{t_{\star} \rho_{h}}{\beta \left( A + 1 \right) \rho_{c}} . 
\end{equation} 
Here $t_{\star}$ is the star-formation timescale in the effective multiphase model, 
$\rho_{h}$ and $\rho_{c}$ are the densities of hot ambient gas and cold clouds respectively, 
$\beta$ is the mass fraction of stars which are short-lived and instantly die as SNe, 
$A$ is the efficiency of evaporation of cold clouds to be returned to the hot phase due to SNe feedback. 
If the local cooing time (computed assuming all the particle mass is in the hot phase) 
is shorter than $\tau_h$, 
normal radiative cooling is used to dissipate the BH thermal feedback energy. 

\noindent $\bullet$ {\bf Kinetic} : The gas velocity is increased, as described next.

\subsection{Kinetic AGN Feedback} 
\label{sec-num-BH-Kin} 

In the following we consider that BH feedback drives a gas outflow 
of velocity $v_w$ and mass outflow rate $\dot{M}_w$. 
The energy-conservation equations can be written using the kinetic energy or momentum 
of the outflowing gas, each of which gives one AGN wind formalism. 
We consider one fixed value for $v_w$ (a free parameter), which is a simplified assumption 
of our models (intended to be applied to cosmological simulations in the future), 
although more physically the AGN wind velocity should be self-regulated 
\citep[e.g.,][]{Gaspari11a, Gaspari11b}.

\subsubsection{Energy-Driven Wind (EDW)} 
\label{sec-num-BH-Kin-EDW} 

The kinetic energy carried away by the wind is equated to the feedback energy from BH: 
\begin{equation} 
\frac{1}{2} \dot{M}_w v_w^2 = \dot{E}_{\rm feed} = \epsilon_f \epsilon_r \dot{M}_{\rm BH} c^2 . 
\end{equation} 
This gives the outflow rate as, 
\begin{equation} 
\label{eq-MdotW-EDW} 
\dot{M}_w = 2 \epsilon_f \epsilon_r \dot{M}_{\rm BH} \frac{c^2}{v_w^2} . 
\end{equation} 

\subsubsection{Momentum-Driven Wind (MDW)} 
\label{sec-num-BH-Kin-MDW} 

Energy output is related to the momentum of radiation via $E = p c$. 
The rate of momentum outflow in the AGN wind is $\dot{p}_w = \dot{M}_w v_w$. 
Equating $\dot{p}_w$ to the radiation momentum from AGN we get, 
\begin{equation} 
\dot{p}_w = \dot{M}_w v_w = \frac{\dot{E}_{\rm feed}}{c} = \epsilon_f \epsilon_r \dot{M}_{\rm BH} c . 
\end{equation} 
This expresses the mass outflow rate in terms of the BH accretion rate, 
\begin{equation} 
\label{eq-MdotW-MDW} 
\dot{M}_w = \epsilon_f \epsilon_r \dot{M}_{\rm BH} \frac{c}{v_w} . 
\end{equation} 

The main difference between EDW and MDW is the occurrence of factors 
$(c / v_w)^2$ and $(c / v_w)$ in the mass outflow rate Eqs.~(\ref{eq-MdotW-EDW}) and (\ref{eq-MdotW-MDW}). 
Hence to have the same $\dot{M}_w$ in both cases, a larger efficiency factor is needed in MDW: 
$\epsilon_{f,{\rm MDW}} = 2 \epsilon_{f,{\rm EDW}} (c / v_w)$. 

There are two free parameters in our subgrid model of kinetic feedback: $\epsilon_f$ and $v_w$. 
Typical AGN wind velocity values seen in observations is between 
$v_w = {\rm a ~ few} ~ 1000 - 10000$ km/s. 
\citet{Debuhr12} considered wind velocities of $3000$, $7000$ and $10000$ km/s 
in their simulations of kinetic AGN feedback. 
The radiative efficiency $\epsilon_r$ (Eqs.~\ref{eq-LEdd}, \ref{eq-Lr-BH}) is held at a fixed value. 
We vary the free parameters ($\epsilon_f, v_w$) within reasonable ranges 
to obtain a closest match of the simulation versus observational 
$[M_{\rm BH} - \sigma_{\star}]$ relation (\S\ref{sec-res-MassBH-SigmaGalaxy}), 
and discuss the best-fit parameters.

\subsection{Implementation in the {\sc GADGET-3} code} 
\label{sec-num-Implement} 

% internal degree of freedom % a subgrid fashion, 

A BH is represented as a collisionless particle in the {\sc GADGET-3} code, 
having a {\it dynamical} mass $m_{\rm BH,dyn}$, 
given by simulation resolution (\S\ref{sec-num-IC}, Table~\ref{Table-Galaxies}). 
Owing to the numerics of low- and medium-resolution simulations 
(where one might need to track BHs in galaxies containing some hundreds of particles), 
there is another variable describing the BH mass in a smooth fashion, 
which we call the {\it subgrid mass}, $M_{\rm BH}$. 
The value $m_{\rm BH,dyn}$ is used for the non-AGN physics in the code (e.g. gravitational interactions). 
$M_{\rm BH}$ is used to compute the AGN physics (e.g. Bondi rate, Eq.~\ref{eq-Mdot-Bondi}), 
which is hence the true BH mass. 

At $t = 0$ in our simulations, the collisionless particle 
is seeded with a BH having an initial subgrid mass $M_{\rm BH} = M_{\rm BH, seed} = 10^5 M_{\odot}$. 
At each timestep $\Delta t$, 
it grows according to the BH accretion rate (Eq.~\ref{eq-Mdot-BH}), 
its subgrid mass increases by an amount $\dot{M}_{\rm BH} \Delta t$, 
with $m_{\rm BH,dyn}$ remaining the same. 
The initial growth from $M_{\rm BH, seed}$ to $m_{\rm BH,dyn}$ occurs in that way, 
without altering the surrounding gas distribution. 
After a BH has grown such that $M_{\rm BH} \geq m_{\rm BH,dyn}$, 
it might accrete (so called {\it swallow}) neighbouring gas particles, using a stochastic methodology. 
When a gas particle is swallowed, it is removed from the simulation, 
and $m_{\rm BH,dyn}$ increases by the swallowed particle mass $m_{\rm gas}$. 
This conserves dynamical mass within the computational volume. 
The probability of swallowing gas is set to ensure that $M_{\rm BH}$ and $m_{\rm BH,dyn}$ 
tracks each other closely. 
Such a procedure grows the BH in a continuous fashion with time, increasing the mass $M_{\rm BH}$ smoothly. 
It also allows to track BHs less massive than $m_{\rm BH,dyn}$. 
Having just a single BH mass would create significant fluctuations in $m_{\rm BH,dyn}$ 
at the epochs when discrete gas particles are accreted, 
and would render impossible to have a correct BH mass in less massive galaxies. 

We do not incorporate any scheme for BH advection 
(which is done in some studies by e.g. reposition BH at minimum gravitational potential), 
since our tests obtain a BH dynamics expected for isolated systems. 
In an isolated galaxy the separation between the BH and the minimum gravitational potential 
is always less than the softening lengths, 
while during a merger the BHs deviate from the potential minima by a few times softening 
some of which is due to merger dynamics. 

% BH is repositioned manually at the min-grav-pot location. 
% Meaning that the BH almost tracks the min-grav-pot in a galaxy undergoing quiescent evolution. 

Kernel-weighted quantities are computed smoothing over gas particles in the local environment around the BH, 
using a kernel having the same shape as that used in SPH calculations. 
However $4$ times more neighbours are used for the BH particle, 
than in the SPH (which has $32 \pm 4$ neighbours). 
The kernel size, or the BH smoothing length $s_{\rm BH}$, is determined 
(in analogy to finding gas particle smoothing length) by implicit solution of the equation, 
\begin{equation} 
\label{eq-BH-Smooth} 
\frac{4}{3} \pi s_{\rm BH}^3 \rho_{\rm BH} = M_{\rm ngb} . 
\end{equation} 
Here $\rho_{\rm BH}$ is the kernel estimate of the gas density at the position of the BH, 
and $M_{\rm ngb}$ is the mass of $\sim 4 \times 32$ neighbouring gas particles. 

We implement a probabilistic criterion (similar to other subgrid prescriptions in {\sc GADGET-3}) 
to distribute the kinetic feedback energy from the BH to the neighbouring gas, i.e. 
particles whose masses contributed to the total neighbour mass $M_{\rm ngb}$ in Eq.~(\ref{eq-BH-Smooth}). 
Gas particles are stochastically selected from the neighbours and kicked into AGN wind, 
by imparting a one-time $v_w$ boost. 
A probability for being kicked is calculated in a timestep $\Delta t$ for each neighbouring $i$'th gas particle: 
\begin{equation} 
p_i = \frac{w_i \dot{M}_w \Delta t} {\rho_{\rm BH}} . 
\end{equation} 
Here $w_i = W( | r_{\rm BH} - r_i |, s_{\rm BH})$ is the SPH kernel weight of the gas particle relative to the BH, 
and $\dot{M}_w$ is the mass outflow rate expressed by 
Eq.~(\ref{eq-MdotW-EDW}) or (\ref{eq-MdotW-MDW}) for EDW or MDW respectively. 
Note that $p_i$ is similar to the probability for swallowing gas particles during BH accretion \citep{SDH05}. 
A random number $x_i$, uniformly distributed in the interval $[0, 1]$, 
is drawn and compared with $p_i$. 
For $x_i < p_i$, the gas particle is given a wind velocity kick. 
If $\vec{v}_{\rm old}$ is particle velocity and $\phi$ its gravitational potential, 
then after receiving AGN wind kick, its velocity is updated to: 
\begin{equation} 
\label{eq-vNew} 
\vec{v}_{\rm new} = \vec{v}_{\rm old} + v_w \hat{x} . 
\end{equation} 
The direction given by the unit vector $\hat{x}$ is set along 
$\left( \vec{v}_{\rm old} \times \vec{\nabla} \phi \right)$ or 
$- \left( \vec{v}_{\rm old} \times \vec{\nabla} \phi \right)$, randomly selected between the two. 
This makes the wind particles to be preferentially ejected along the 
rotation axis of the galaxy or perpendicular to the galaxy disk. 
Some other studies \citep[e.g.,][]{Tescari11, Barai13} implement hydrodynamic decoupling of the wind particles. 
%to enable the outflow to escape without affecting star-formation in the galaxy. 
Unlike those we do not allow any decoupling, 
i.e. in this work the AGN wind particles are always coupled and undergo hydrodynamic interactions. 

The merger criterion of two BHs is when they come inside the $s_{\rm BH}$ of each other, 
and their relative velocity is smaller than the local sound speed.

\subsubsection{Detect Hole in Gas Distribution, and Prevent its Growth} 
\label{sec-num-HolePrevent} 

We implement a novel numerical algorithm in the {\sc GADGET-3} code 
to detect and prevent the expansion of hole in the gas distribution around the BH. 
In the original version, there was a problem of hole creation, 
classically demonstrated by a BH accreting gas 
with no-feedback at the center of a rotationally supported disk galaxy. 
The BH would deplete the central gas, inside its smoothing length, by swallowing particles. 
In order to have a constant number of neighbours, 
$s_{\rm BH}$ increases after some time encompassing gas further out. 
The new gas is depleted in turn, turning the BH more massive, and 
creating an enlarging hole at the galaxy center. 
The process continues ad infinitum, 
with $s_{\rm BH}$ extending furthermore to accrete the gas of the whole galaxy, 
unless an upper limit is imposed on $s_{\rm BH}$. 
Here gas is artificially accreted because of the numerical scheme, 
and not physically because it has flown inward. 
Observed galaxies do not show any hole in the gas distribution around their central BHs, 
therefore the creation of such artificial holes affects 
the simulated evolution of the galaxy and AGN feedback in unwanted ways. 

This issue has been present in all the studies using the BH numerical methodology of \citet{SDH05}, 
demonstrated by the BH growing to $M_{\rm BH} > 10^9 M_{\odot}$ (their Fig.~10) in a no-feedback run. 
A visual example of the hole can be seen in Fig.~13 of \citet{Wurster13b}. 

One solution is to set a maximum $s_{\rm BH}$ manually, 
which however is not elegant because the value would vary with resolution, galaxy mass, 
and additionally the environment in a cosmological simulation. 
We have rather worked out a computational solution, 
to prevent the limitless increase of $s_{\rm BH}$ independent of simulation conditions. 

Our numerical methodology assumes that the BH lies at the gas density peak, 
or minimum SPH smoothing length ($s_{\rm sml}$) location. 
The existence of a {\it hole} around the BH is detected using the $s_{\rm sml}$ distribution of neighbours. 
With no hole, the minimum $s_{\rm sml}$ occurs at the BH position. 
When there is a hole, a preferential boundary is created at $s_{\rm BH}$, 
with a fewer than expected number of gas particles inside. 
This causes a small increase in the $s_{\rm sml}$ of neighbours nearest to the BH, 
and the minimum $s_{\rm sml}$ occurs at a finite distance from the BH location. 
Also $s_{\rm BH}$ is then more than $2$ times larger than the minimum $s_{\rm sml}$ of neighbours. 

In our code implementation, 
we detect just when a hole is created around a BH, and control it. 
Gas particles lying within a multiplication factor $d_h$ times $s_{\rm BH}$ are searched 
to find the nearest particle's smoothing length $s_{\rm near}$, 
and the minimum smoothing length $s_{\rm min}$. 
If either of the following ratio of smoothing lengths exceeds a value $r_h$: 
\begin{equation} 
\label{eq-Hole-Detect} 
\frac{s_{\rm BH}}{s_{\rm min}} > r_h, ~~~~~~~~~~ {\rm or}, ~~~~~~~~~~ \frac{s_{\rm near}}{s_{\rm min}} > r_h , 
\end{equation} 
then a hole exists. 
When this {\it existence-of-hole} condition is met for a BH, its $s_{\rm BH}$ is held fixed, 
and not allowed to enlarge further. 
Testing with a single BH in an isolated galaxy we find the working values of factors as: 
$d_h = 4$, and $r_h = 1.7$. 
The hole is limited to a size $\sim (0.7 - 0.8) h^{-1}$ kpc successfully. 
All our isolated and merger simulations are done using these values in the hole-detection algorithm. 

% PB_BH_HSML_PREVENT_HOLE             # Prevent expansion of Hole around BH using h_sml pattern of gas neighbours. 
% PB_BH_CHECK_HOLE_FIND_NGB_DIST=4.0  # Multiplication factor of BH h_sml to search neighbours for hole detection. 
% PB_BH_CHECK_HOLE_RATIO_LIMIT=1.7    # Smoothing Length Ratio to check existence of Hole. 
% 
%  h_i  = PB_BH_CHECK_HOLE_FIND_NGB_DIST * h_i; /* Limiting distance to find neighbours = factor * BH h_sml. */ 
% 
%	   if( (Ratio_Hsml_BH_by_Min > PB_BH_CHECK_HOLE_RATIO_LIMIT) || (Ratio_Hsml_AtrMin_by_Min > PB_BH_CHECK_HOLE_RATIO_LIMIT) ) 
%	   { 
%	      if(PPP[n].Hsml < All.BlackHoleMaxAccretionRadius) 
%		 BH_MaxAccretionRadius = PPP[n].Hsml; 
%	      else 
%		 BH_MaxAccretionRadius = All.BlackHoleMaxAccretionRadius; 
%	   } 
%	   else 
%	   { 
%	      if(Ratio_r_AThsmlMin_by_Min < PB_BH_CHECK_HOLE_RATIO_LIMIT) 
%		 BH_MaxAccretionRadius = All.BlackHoleMaxAccretionRadius; 
%	   } 
% 
% Computationally this new hole-detection algorithm in the code 
% requires to perform an additional neighbour search for each BH. 
% The current BH accretion routine already does 2 neighbour searches, makes it a total of 3 neighbour searches. 
% Keeping the HOLE fixed to a small-radius from just when it starts to appear  

\subsection{Initial Galaxy Models} 
\label{sec-num-IC} 

The initial isolated galaxy models are constructed following the approach 
described in \citet{Springel99, SDH05}. 
The total galaxy mass $M_{\rm tot} = v_{200}^3 / \left( 10 G H_{0} \right)$, 
is expressed in terms of the circular virial velocity $v_{200}$. 
A Hubble parameter of $H_{0} = 70.3$ km s$^{-1}$ Mpc$^{-1}$, 
or $h = 0.703$, \citep[e.g.][]{Komatsu11} is adopted. 
Each galaxy is composed of a dark matter (DM) halo, 
a rotationally supported gaseous and stellar disk comprising of a fraction $f_{d} = 0.04$ of the total mass, 
and a central stellar bulge of mass fraction $f_{b} = 0.01$. 
The mass distribution of the DM halo is modelled with the \citet{Hernquist90} profile, 
and has a spin parameter of $\lambda = 0.04$. 
The disk has a mass fraction $f_{\rm gas} = 0.2$ as gas, and the rest as stars; 
both components are modelled with an exponential surface density profile 
of radial scalelength $d$, 
and radially-constant vertical scaleheight $z_{0} = 0.2 d$. 
The spherical stellar bulge is modelled with a \citet{Hernquist90} profile having a scalelength $b = 0.2 d$. 

%%%%%%%%%%%%%%%%%%%%%%%%%%%%%%%%%%%%%%%%%%%%%%%%%%%%%%%%%%%%%%%%%%%%%%%%%%%%%%%%%%%%%%%%%%
%%%%%%%%%%%%%%%%%%%%%%%%%%%%%%%%%%%%%%%%%%%%%%%%%%%%%%%%%%%%%%%%%%%%%%%%%%%%%%%%%%%%%%%%%% 
%
% TABLE 1 

\begin{table*} 
%\begin{sidewaystable} 
%\centering 
\scriptsize 
\begin{minipage}{0.95 \linewidth} 
\caption{ 
Galaxy Initial Conditions. 
Column 2: Virial velocity. 
Column 3: Total mass (dark matter + gas + stars). 
Column 4: Gas mass (disk). 
Column 5: Stellar disk mass. 
Column 6: Stellar bulge mass. 
Column 7: Dark matter particle mass. 
Column 8: Gas particle mass. 
Column 9: Disk star particle mass. 
Column 10: BH particle dynamical mass. 
Column 11: Disk (gas + stars) scale length. 
} 
\label{Table-Galaxies} 
\begin{tabular}{@{}ccccccccccc} 

\hline 

Series & $v_{200}$ & $M_{\rm tot}$ & $M_{\rm gas}$ & $M_{{\star},{\rm disk}}$ & $M_{{\star},{\rm bulge}}$ & $m_{\rm DM}$ & $m_{\rm gas}$ & $m_{{\star},{\rm disk}}$ & $m_{\rm BH,dyn}$ & $d$ \\ 

Name & (km/s) & [$M_{\odot}$] & [$M_{\odot}$] & [$M_{\odot}$] & [$M_{\odot}$] & [$M_{\odot}$] & [$M_{\odot}$] & [$M_{\odot}$] & [$M_{\odot}$] & (kpc/h) \\ 

\hline 

Low-mass & $75$ & $1.40 \times 10^{11}$ & $1.12 \times 10^9$ & $4.47 \times 10^9$ & $1.40 \times 10^9$ & $4.42 \times 10^5$ & $2.23 \times 10^4$ & $1.79 \times 10^5$ & $6.98 \times 10^5$ & $1.49$ \\ 

Fiducial & $150$ & $1.12 \times 10^{12}$ & $8.93 \times 10^9$ & $3.57 \times 10^{10}$ & $1.12 \times 10^{10}$ & $3.53 \times 10^6$ & $1.79 \times 10^5$ & $1.43 \times 10^6$ & $5.58 \times 10^6$ & $2.99$ \\ 

High-mass & $300$ & $8.93 \times 10^{12}$ & $7.15 \times 10^{10}$ & $2.86 \times 10^{11}$ & $8.93 \times 10^{10}$ & $2.83 \times 10^7$ & $1.43 \times 10^6$ & $1.14 \times 10^7$ & $4.47 \times 10^7$ & $5.98$ \\ 

\hline 

\end{tabular} 
\end{minipage} 
%\end{sidewaystable} 
\end{table*} 

%%%%%%%%%%%%%%%%%%%%%%%%%%%%%%%%%%%%%%%%%%%%%%%%%%%%%%%%%%%%%%%%%%%%%%%%%%%%%%%%%%%%%%%%%% 
%%%%%%%%%%%%%%%%%%%%%%%%%%%%%%%%%%%%%%%%%%%%%%%%%%%%%%%%%%%%%%%%%%%%%%%%%%%%%%%%%%%%%%%%%% 

We simulated galaxies of three masses. 
Our fiducial galaxy is generated using $v_{200} = 150$ km/s, 
which corresponds to $M_{\rm tot} = 1.12 \times 10^{12} M_{\odot}$, a similar mass as the Milky-Way. 
Furthermore, we simulate galaxies having a lower-mass with $v_{200} = 75$ km/s, 
and a higher-mass with $v_{200} = 300$ km/s. 
The number of particles of various types in the initial condition of each galaxy are: 
$3 \times 10^5$ DM, $5 \times 10^4$ gas, $25 \times 10^3$ disk stars, and $25 \times 10^3$ bulge stars. 
Table~\ref{Table-Galaxies} lists the relevant mass components and particle masses of all the galaxies. 
New star particles form during the simulation (via star formation in the gas), 
which are less massive than the initial stellar particles in the disk and bulge. 
All the particles (DM, gas, stars) follow collisionless gravitational dynamics, 
while in addition the gas particles undergo hydrodynamical interactions. 

A collisionless tracer particle of mass fraction $f_{\rm BH} = 5 \times 10^{-6}$ 
is generated at the centre of a galaxy to {\it carry} the SMBH. 
This corresponds to a dynamical BH particle mass ($5.58 \times 10^6 M_{\odot}$ in our fiducial galaxy) 
which is $\sim 1.6$ times higher than the DM particle mass. 
The BH particle is thus expected to trace the minimum of the gravitational potential closely, 
minimizing artificial dynamical motion. 
In the AGN simulations, 
a BH of initial subgrid mass $10^5 M_{\odot}$ is seeded in this tracer particle. 

The Plummer-equivalent softening length for gravitational forces is set to 
$L_{\rm soft} = 0.5 / h$ kpc for the gas and star particles, and $1 / h$ kpc for the DM and BH particles. 
The minimum SPH smoothing length is set to a fraction $0.001$ of $L_{\rm soft}$. 

As the initial conditions for the merger simulations, 
two equal-mass isolated galaxies are generated using a fixed $v_{200}$. 
The orbital planes of the two disks are kept the same, and 
they are set on a parabolic collision course in the disk plane. 
A minimum separation at which the galaxies would pass at pericenter (if they were point masses) 
is taken as $2.5 / h$, $5 / h$, and $10 / h$ kpc, 
respectively for the lower-mass, fiducial, and higher-mass cases.

\subsection{Simulations} 
\label{sec-num-Sim} 

%%%%%%%%%%%%%%%%%%%%%%%%%%%%%%%%%%%%%%%%%%%%%%%%%%%%%%%%%%%%%%%%%%%%%%%%%%%%%%%%%%%%%%%%%% 
%%%%%%%%%%%%%%%%%%%%%%%%%%%%%%%%%%%%%%%%%%%%%%%%%%%%%%%%%%%%%%%%%%%%%%%%%%%%%%%%%%%%%%%%%% 
% 
% TABLE 2 

\begin{table*} 
\begin{minipage}{0.8 \linewidth} 
\caption{ 
Simulation Parameters. 
Column 1: Name of simulation run. 
Column 2: Feedback efficiency, $\epsilon_f$ = 
Fraction of the radiated energy from BH which is coupled to the surrounding gas. 
Column 3: $v_w$ = Outflow velocity in kinetic feedback prescription. 
Column 4: Specifications of AGN feedback model. 
Columns 5, 6, 7: Letters 'I' (Isolated galaxy evolution) and/or 'M' (galaxy Merger) 
is written below Lower-Mass ($v_{200} = 75$ km/s), 
Fiducial ($v_{200} = 150$ km/s), and Higher-Mass ($v_{200} = 300$ km/s) galaxy models, 
wherever the parameter set in the row has been run with the column initial configuration. 
} 
\label{Table-Sims} 
\begin{tabular}{@{}ccccccc} 

\hline 

Run & $\epsilon_f$ & $v_w$ & AGN Feedback & Lower-Mass & Fiducial & Higher-Mass \\ 
Name & & [km/s] & \\ 

\hline 

{\it SF} & & & No BH & & I, M & \\ 

\hline 

{\it th1} & $0.002$ & & BH Thermal &   & I, M &   \\ 
{\it th2} & $0.01$  & & BH Thermal & I & I, M & I \\ 
{\it th3} & $0.05$  & & BH Thermal & M & I, M & M \\ 

\hline 

{\it kinE1} & $0.01$ & $5,000$ & BH Kinetic: Energy-driven wind &   & I, M &   \\ 
{\it kinE2} & $0.05$ & $5,000$ & BH Kinetic: Energy-driven wind & I & I, M & I \\ 
{\it kinE3} & $0.25$ & $5,000$ & BH Kinetic: Energy-driven wind & M & I, M & M \\ 

{\it kinE4} & $0.01$ & $10,000$ & BH Kinetic: Energy-driven wind &      & I, M &      \\ 
{\it kinE5} & $0.05$ & $10,000$ & BH Kinetic: Energy-driven wind &      & I, M &      \\ 
{\it kinE6} & $0.25$ & $10,000$ & BH Kinetic: Energy-driven wind & I, M & I, M & I, M \\ 

\hline 

{\it kinM1} & $0.25$ & $5,000$ & BH Kinetic: Momentum-driven wind &   & I, M &   \\ 
{\it kinM2} & $1.0$  & $5,000$ & BH Kinetic: Momentum-driven wind &   & I, M &   \\ 
{\it kinM3} & $1.0$  & $2,500$ & BH Kinetic: Momentum-driven wind & I & I, M & I \\ 
{\it kinM4} & $1.0$  & $1,000$ & BH Kinetic: Momentum-driven wind &   &    M &   \\ 

\hline 
\end{tabular} 

\end{minipage} 
\end{table*} 

%%%%%%%%%%%%%%%%%%%%%%%%%%%%%%%%%%%%%%%%%%%%%%%%%%%%%%%%%%%%%%%%%%%%%%%%%%%%%%%%%%%%%%%%%% 
%%%%%%%%%%%%%%%%%%%%%%%%%%%%%%%%%%%%%%%%%%%%%%%%%%%%%%%%%%%%%%%%%%%%%%%%%%%%%%%%%%%%%%%%%% 

Table~\ref{Table-Sims} lists the series of simulations we perform. 
The different runs incorporate the same non-AGN subgrid physics described 
in \S\ref{sec-numerical}, and investigate different AGN feedback models. 
They are chosen from four broad categories, as given below, exploring the model parameter space. 

\noindent $\bullet$ {\it SF} : 
One run with star-formation, stellar evolution, and chemical enrichment only (no BH). 

\noindent $\bullet$ {\it th1 --- th3} : 
Three runs of thermal feedback from BH (\S\ref{sec-num-BH-Accr-Feed}), 
using $\epsilon_f = 0.002, 0.01, 0.05$. 

\noindent $\bullet$ {\it kinE1 --- kinE6} : 
Six runs of kinetic feedback from BH with energy-driven wind prescription (\S\ref{sec-num-BH-Kin-EDW}), 
using combinations of $\epsilon_f = 0.01, 0.05, 0.25$, and $v_w = 5000, 10000$ km/s. 

\noindent $\bullet$ {\it kinM1 --- kinM4} : 
Four runs of kinetic feedback from BH with momentum-driven wind prescription 
(\S\ref{sec-num-BH-Kin-MDW}), 
using combinations of $\epsilon_f = 0.25, 1.0$, and $v_w = 1000, 2500, 5000$ km/s. 

The fiducial galaxy ($v_{200} = 150$ km/s) is simulated using all the above parameter cases 
for both isolated galaxy evolution (indicated by letter 'I' in Table~\ref{Table-Sims}, Columns 5, 6, 7), 
and galaxy merger (indicated by letter 'M' in Table~\ref{Table-Sims}). 
The parameter set of each category which generates a closest fit to observations 
is also run with the lower-mass ($v_{200} = 75$ km/s, and higher-mass ($v_{200} = 300$ km/s) galaxy models. 
The isolated galaxies are evolved up to a time $2$ Gyr, and the galaxy mergers up to $3$ Gyr.

\section{Results and Analysis}
\label{sec-results} 

We compare the black hole mass $(M_{\rm BH})$ versus 
galaxy stellar velocity dispersion $(\sigma_{\star})$ results obtained in our simulations 
with the observed correlation, using a manual chi-by-eye approach. 
This comparison is considered as the figure-of-merit 
in order to obtain the best-fit parameters for each AGN feedback model. 
The fiducial galaxy ($v_{200} = 150$ km/s) is simulated using several parameter variations 
in a series of runs (\S\ref{sec-num-Sim}), each for isolated and merger evolution. 
% for thermal BH feedback $\epsilon_f = 0.002, 0.01, 0.05$; 
% for kinetic BH feedback with EDW prescription combinations of 
% $\epsilon_f = 0.01, 0.05, 0.25$ and $v_w = 5000, 10000$ km/s; 
% for kinetic BH feedback with MDW prescription combinations of 
% $\epsilon_f = 0.25, 1$ and $v_w = 1000, 2500, 5000$ km/s. 
The parameter set of each category which generates a closest fit to the observational 
$[M_{\rm BH} - \sigma_{\star}]$ relation at the fiducial galaxy mass are selected. 
These chosen cases are also run with the 
lower-mass ($v_{200} = 75$ km/s), and higher-mass ($v_{200} = 300$ km/s) galaxy models. 

All the runs performed are presented in \S\ref{sec-res-MassBH-SigmaGalaxy}. 
For the galaxy morphologies in \S\ref{sec-res-Morph-Outflow}, 
we choose models generating a comparable BH mass, 
i.e. the BHs in these cases exist in the same region of the $[M_{\rm BH} - \sigma_{\star}]$ diagram. 
Consequently we compare between thermal and kinetic feedback with different $\epsilon_f$, 
because the respective best-fit $\epsilon_f$ are unequal. 
Results in remaining sections only show the closest fit models, 
plus a few cases with varying $\epsilon_f$ and $v_w$. 
Note that our approach explores parameters which fit the $[M_{\rm BH} - \sigma_{\star}]$ relation. 
This however renders difficult to perform an absolute comparison of thermal versus kinetic feedback, 
since the respective output powers are different, as described in \S\ref{sec-res-BHAR-SFR}. 

% but we can not compare 1:1 both types feedback for other matters (e.g. profiles, outflowing mass, etc.). 
% to judge between different parameter values % best fitting the observations 
% where we select the best-fit parameters. 

We analyse the carbon content of the gas in the galaxies, 
since it is one of the most abundant heavy element in the Universe, 
and the spectral lines produced by ionized carbon are relatively easy to observe. 
The carbon metallicity, $Z_C$, 
is computed as the ratio of carbon mass to the total particle mass for each gas particle. 
Abundance ratios are expressed in terms of the Solar metallicity, 
which is $Z_{C, \odot} = 0.002177$ (mass fraction of carbon in Sun) 
derived from the compilation by \citet{Asplund05}.

\subsection{$[M_{\rm BH} - \sigma_{\star}]$ Correlation} 
\label{sec-res-MassBH-SigmaGalaxy} 

%%%%%%%%%%%%%%%%%%%%%%%%%%%%%%%%%%%%%%%%%%%%%%%%%%%%%%%%%%%%%%%%%%%%%%%% 

% FIGURE 1 
\begin{figure*} 
\centering 
\includegraphics[width = 1.0 \linewidth]{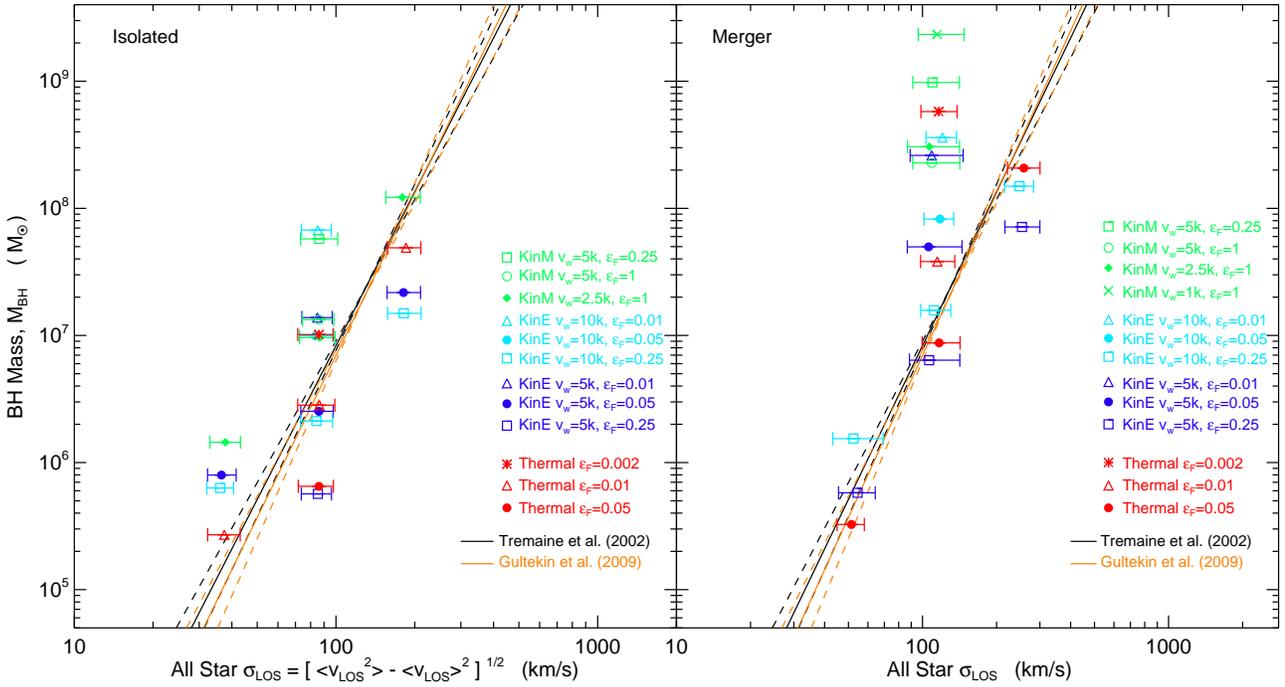} 
\caption{ 
Black hole mass versus galaxy stellar velocity dispersion along line-of-sight. 
The single isolated galaxy models, shown at an evolution time of $2$ Gyr, are in the {\it left} panel. 
The merger simulations, shown at an evolution time of $2.5$ Gyr, are in the {\it right} panel. 
The different colours and plotting symbols distinguish the AGN models as labelled in each panel. 
Each of the four broad categories of feedback are denoted by a different colour: 
[{\it th1 --- th3}] thermal - {\it red}, 
[{\it kinE1 --- kinE3}] kinetic EDW with $v_w =  5000$ km/s - {\it blue}, 
[{\it kinE4 --- kinE6}] kinetic EDW with $v_w = 10000$ km/s - {\it cyan}, 
[{\it kinM1 --- kinM4}] kinetic MDW - {\it green}. 
The parameter choices are represented by the plotting symbols: 
$\epsilon_f = 0.002$ - {\it asterisk}, 
$\epsilon_f = 0.01$ - {\it triangle}, 
$\epsilon_f = 0.05$ - {\it filled circle}, 
$\epsilon_f = 0.25$ - {\it open square}, 
$\epsilon_f = 1$ - {\it open circle}, 
$\epsilon_f = 1$ and $v_w =  2500$ km/s - {\it filled diamond}, 
$\epsilon_f = 1$ and $v_w =  1000$ km/s - {\it cross}. 
The solid and dashed lines display the best-fitting relations and error bars obtained from observations 
by \citet{Tremaine02} - black, and \citet{Gultekin09} - orange. 
} 
\label{fig-MassBH-SigmaStarLOS} 
\end{figure*} 

%%%%%%%%%%%%%%%%%%%%%%%%%%%%%%%%%%%%%%%%%%%%%%%%%%%%%%%%%%%%%%%%%%%%%%%% 

The galaxy stellar velocity dispersion $\sigma_{\star}$ is computed by considering all the stars: 
those newly formed during the simulation evolution by SF from gas, 
the disk and bulge stellar components present from the initial condition 
(\S\ref{sec-num-IC}, Table~\ref{Table-Galaxies}). 
The position of the BH is taken as the galaxy center, 
and the distances (or radii) of star particles are estimated from it. 
In the case of merging galaxies the calculations are done at a time when the BHs have merged 
to produce a single BH, defining an unique galaxy center. 
The cumulative sum of all star's mass versus radius is found, 
as well as the radius $R_{1/2}$ containing $1/2$ of the total stellar mass. 
This stellar half-mass radius is considered as the effective radius in estimating $\sigma_{\star}$. 
One hundred random line-of-sight (LOS) directions are chosen around the center 
(or, BH position). 
All the stars lying within $R_{1/2}$ from the center are picked, 
and the LOS velocity component of each, $v_{\rm LOS}$, is found. 
The stellar velocity dispersion along each LOS direction is computed by summing over all relevant stars: 
\begin{equation} 
\label{eq-sigmaLOS} 
\sigma_{\star} = \left( \langle v_{\rm LOS}^2 \rangle - \langle v_{\rm LOS} \rangle^2 \right)^{1/2}. 
\end{equation} 
The same is done for the $100$ LOS directions. 
The median and percentile of the $100$ random direction $\sigma_{\star}$ values are found, 
and shown as our results. 

Fig.~\ref{fig-MassBH-SigmaStarLOS} presents the $M_{\rm BH}$ versus $\sigma_{\star}$ diagram 
obtained in our simulations. 
The {\it left} panel shows the isolated galaxy cases at an evolution time $2$ Gyr. 
The {\it right} panel shows the galaxy merger runs at a time $2.5$ Gyr, 
an epoch by when the two BHs merge in all the cases. 
% The AGN feedback models are differentiated by the colours and plotting symbols, as labelled in each panel. 
The median $\sigma_{\star}$ is depicted by the plotting symbol, 
and 70th percentiles around the median indicated by the lower and upper error bars. 
Observational data is overplotted in Fig.~\ref{fig-MassBH-SigmaStarLOS} as the straight lines. 
The solid and dashed lines display the best-fitting relations and error bars obtained by 
\citet{Tremaine02} in black, and \citet{Gultekin09} in orange. 

The best-fit parameters that we obtain for the isolated galaxy evolution are: 
for BH thermal feedback $\epsilon_f = 0.01$; 
for BH kinetic EDW feedback 
$v_w = 5000$ km/s with $\epsilon_f = 0.05$, and $v_w = 10000$ km/s with $\epsilon_f = 0.25$; 
for BH kinetic MDW feedback $v_w = 2500$ km/s with $\epsilon_f = 1$. 
% These are denoted in Fig.~\ref{fig-MassBH-SigmaStarLOS} left panel 
% by the red triangle, blue filled circle, cyan square, and green filled diamond. 
Estimating by-eye the nearness of the simulation result 
to the observational $[M_{\rm BH} - \sigma_{\star}]$ relation at a given galaxy mass, 
we find mass dependence of the relative fit given by a set of parameters, 
more prominently for the isolated galaxy than the merger case. 
The final BH masses of the 4 best-fit cases 
tend to be relatively smaller than the observations for the higher-mass galaxy, 
and larger than the observations for the lower-mass galaxy. 

We obtain the following best-fit parameters for the galaxy merger: 
for BH thermal feedback $\epsilon_f = 0.05$; 
for BH kinetic EDW feedback $v_w = 5000$ or $10000$ km/s with $\epsilon_f = 0.25$. 
% These are denoted in Fig.~\ref{fig-MassBH-SigmaStarLOS} right panel 
% by the red filled circle, blue and cyan squares. 
For BH kinetic MDW feedback, none of the parameters we explored fit the observations; 
the BH mass is always too large. 
In the case of other feedback models, 
$M_{\rm BH}$ reduces when $\epsilon_f$ is increased (thermal, kinetic EDW), 
also when $v_w$ is decreased (kinetic EDW). 
However there is a reversal of trends with kinetic MDW;   
at $v_w = 5000$ km/s $M_{\rm BH}$ decreases as the efficiency rises from $\epsilon_f = 0.25$ to $1$, 
but $M_{\rm BH}$ increases rather as $v_w$ is reduced further to $2500$ km/s. 
Even when $\epsilon_f$ is set to $1$, 
the smallest BH mass produced is $M_{\rm BH} \sim 2.5 \times 10^8 M_{\odot}$ 
with $v_w = 5000$ km/s in a fiducial galaxy run. 
With $v_w = 1000$ km/s the BH grows drastically to $M_{\rm BH} \sim 2 \times 10^9 M_{\odot}$. 
This happens because in the merging system such a velocity is not high enough to remove the gas away, 
but the kicked gas falls back near the BH(s) and is accreted. 
Thus we find that it is not possible to find $\epsilon_f$ and $v_w$ values to fit 
the observational $[M_{\rm BH} - \sigma_{\star}]$ relation with 
momentum-driven wind prescription of BH kinetic feedback in the case of a galaxy merger.

\subsection{Galaxy Morphology and Outflow} 
\label{sec-res-Morph-Outflow} 

%%%%%%%%%%%%%%%%%%%%%%%%%%%%%%%%%%%%%%%%%%%%%%%%%%%%%%%%%%%%%%%%%%%%%%%% 

% FIGURE 2 
\begin{figure*} 
\centering 
\includegraphics[width = 1.05 \linewidth]{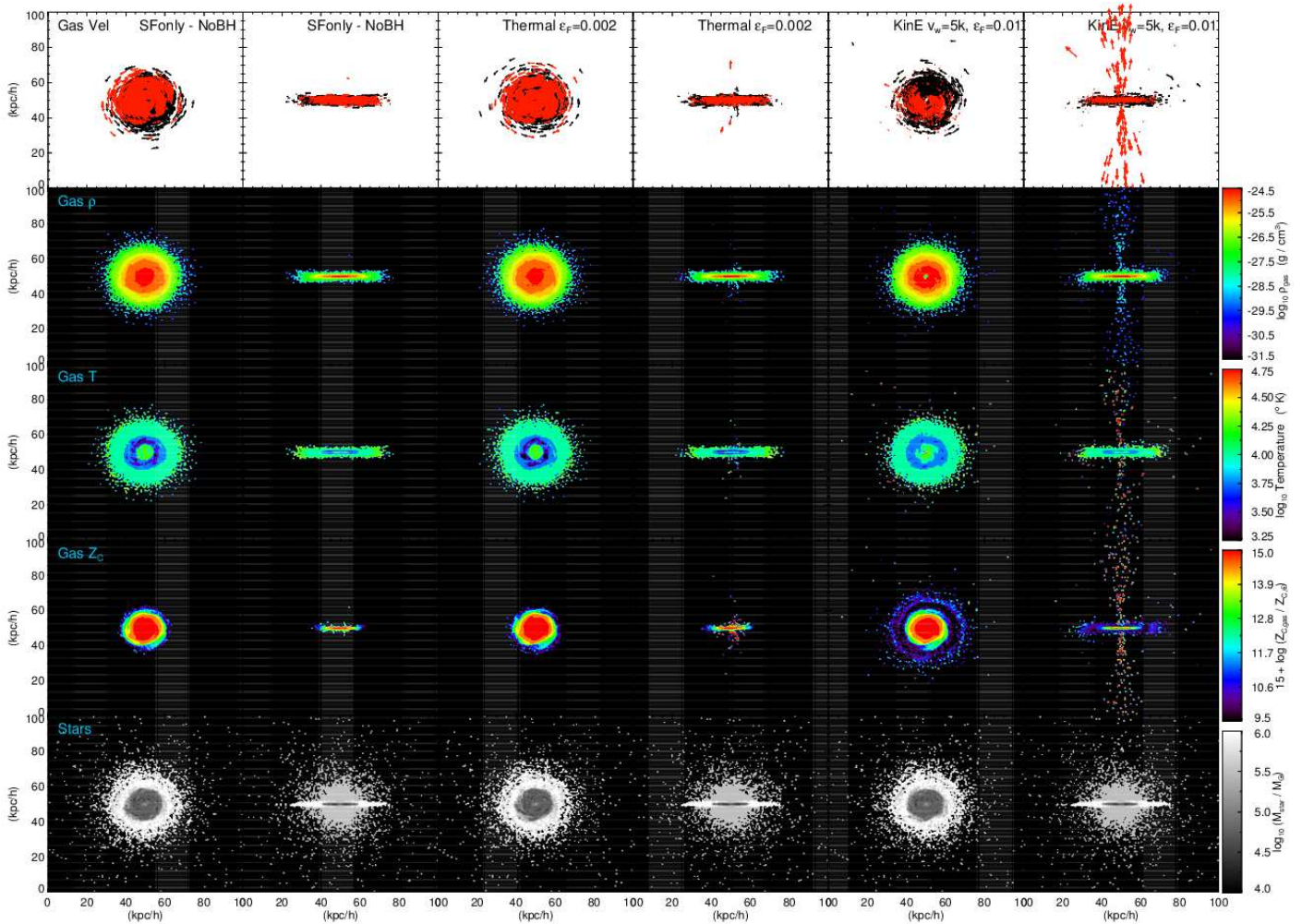} 
\caption{ 
Projection of gas kinematics (top four rows) and stars (bottom row) in isolated fiducial galaxy simulation 
with different feedback models: 
{\it SF} (left 2 columns) - star formation only, 
{\it th1} (middle 2 columns) - BH thermal feedback with $\epsilon_f = 0.002$, 
and {\it kinE1} (right 2 columns) - BH kinetic feedback EDW with $\epsilon_f = 0.01$, $v_w = 5000$ km/s. 
The panels show the face-on (left) and edge-on (right) planes of a $(100 h^{-1}$ kpc$)^3$ volume 
centered around the BH (or collisionless tracer particle in the {\it SF} run) 
at time $t = 1.73$ Gyr. % snap_062 
First row depicts the velocity vectors of $20 \%$ of all the gas particles within the projected volume, 
with the outflowing ($v_r > 0$) particles denoted as red, and the inflowing ($v_r < 0$) as black. 
Second row shows gas density, third row is gas temperature, and fourth row is gas carbon metallicity, 
all projected values, colour coded from red as the highest and black as the lowest. 
Bottom row presents the projected stellar mass, counting all stars (disk + bulge + newly formed in simulation). 
} 
\label{fig-XY-YZ-Isolated} 
\end{figure*} 

%%%%%%%%%%%%%%%%%%%%%%%%%%%%%%%%%%%%%%%%%%%%%%%%%%%%%%%%%%%%%%%%%%%%%%%% 

The galaxy morphology and outflow structure of three representative isolated cases 
are plotted in Fig.~\ref{fig-XY-YZ-Isolated} at an evolution time $1.73$ Gyr. 
It is a fiducial galaxy % undergoing isolated evolution 
with different feedback models: 
{\it SF} (left two columns) - star formation only, 
{\it th1} (middle two columns) - BH thermal feedback with $\epsilon_f = 0.002$, 
and {\it kinE1} (right two columns) - BH kinetic feedback EDW with $\epsilon_f = 0.01$, $v_w = 5000$ km/s. 
In the latter two runs the BHs grow to a comparable mass 
$M_{\rm BH} \sim 1.5 \times 10^7 M_{\odot}$ (from Fig.~\ref{fig-MassBH-SigmaStarLOS}). 

%, at an evolution time $2$ Gyr. 
% Each of the top four rows in Fig.~\ref{fig-XY-YZ-Isolated} shows a gas property, 
% and the bottom row reveals the stars. 
% The BH (or, collisionless tracer particle in the {\it SF} run) position is considered as the center, 
% The panels show the projected quantities in the face-on and edge-on planes 
% of a surrounding $(100 h^{-1}$ kpc$)^3$ volume. 
% In the first row, the outflowing (radial velocity with respect to the BH position, $v_r > 0$) 
% particles are denoted as red, and the inflowing ($v_r < 0$) as black. 
% This row depicts the velocity vectors of $10 \%$ outflowing gas particles and that $10 \%$ inflowing. 

The gas disk of the galaxy retains its identity in all the runs, 
visible as a well-defined rotating disk in the central $r = (10 - 20) h^{-1}$ kpc regions. 
There is no outflow in the {\it SF} case. 
Thermal feedback (run {\it th1}) produces a weak outflow with some gas going out to $(30 - 40) h^{-1}$ kpc, 
but later in time most of the outflowing gas fall back to the disk. 
Kinetic feedback (edge-on plane of {\it kinE1}) produces a well-developed bipolar gas outflow 
propagating perpendicular to the galaxy disk, escaping to $r > 100$ kpc from the central BH position, 
seen in the topmost-right panel, as the red arrows upward and downward directed. 

The gas density, temperature, and carbon metallicity are plotted 
in the second, third, and fourth rows of Fig.~\ref{fig-XY-YZ-Isolated} respectively. 
All the runs have a central overdense region, 
the outer half of which correspond to a cold, annular ring 
composed of gas cooling in the disk, on the way to SF. 
There is a large central concentration of metals, originating from SF, in all the runs 
The metallicity distribution is more centrally concentrated in {\it SF} and {\it th1} cases. 
Kinetic feedback carries some metals out from the SF regions and enrich the CGM and IGM to $> 100$ kpc. 
A remarkable spiral pattern is visible in the face-on $Z_C$ distribution in the 4th row, 5th column panel. 

% colour coded on a log-scale from red as the highest and black as the lowest values. 
% suggesting that kinetic feedback EDW might promote the formation of star-forming gas spiral arms. 

The bottom row of Fig.~\ref{fig-XY-YZ-Isolated} depicts the projected stellar mass, where 
all stars (disk, bulge, newly formed in simulation from gas particles by active SF) have been counted. 
The edge-on plane shows the co-existence of a disk-like and a bulge-like stellar components. 
The star distribution is indistinguishable between the three runs.

%%%%%%%%%%%%%%%%%%%%%%%%%%%%%%%%%%%%%%%%%%%%%%%%%%%%%%%%%%%%%%%%%%%%%%%% 

% FIGURE 3 
\begin{figure*} 
\centering 
$ 
\begin{array}{c} 
\includegraphics[width = 1.05 \linewidth]{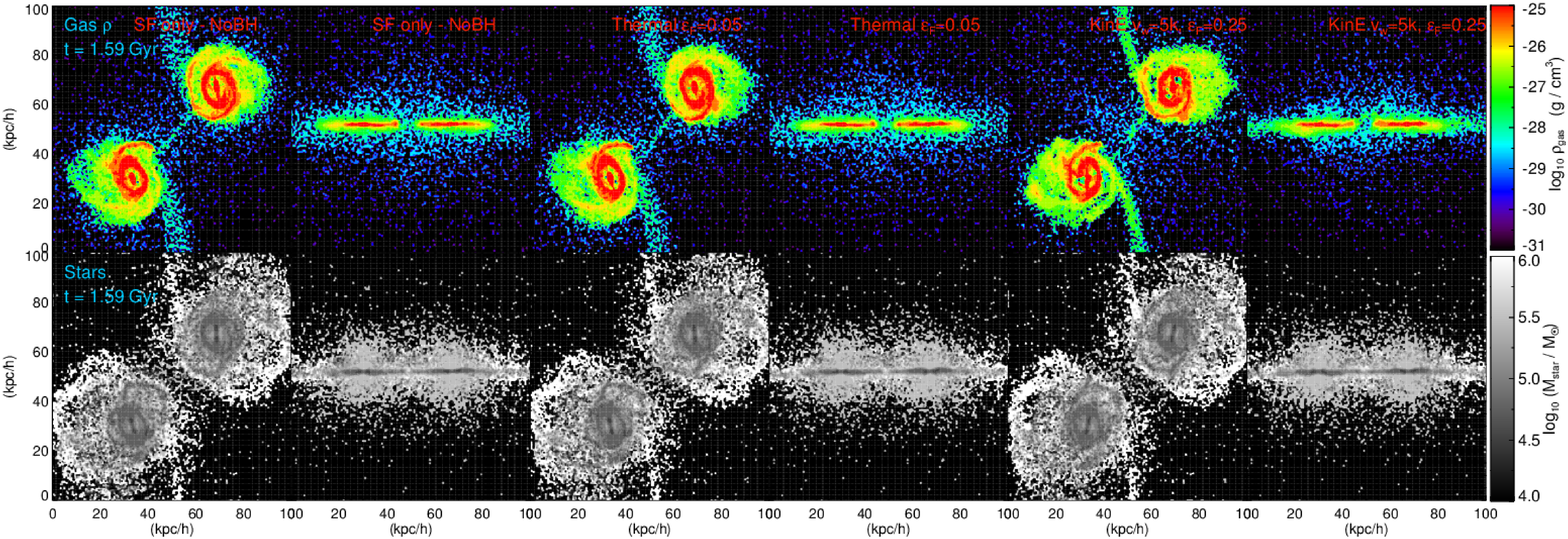} \\ 
\includegraphics[width = 1.05 \linewidth]{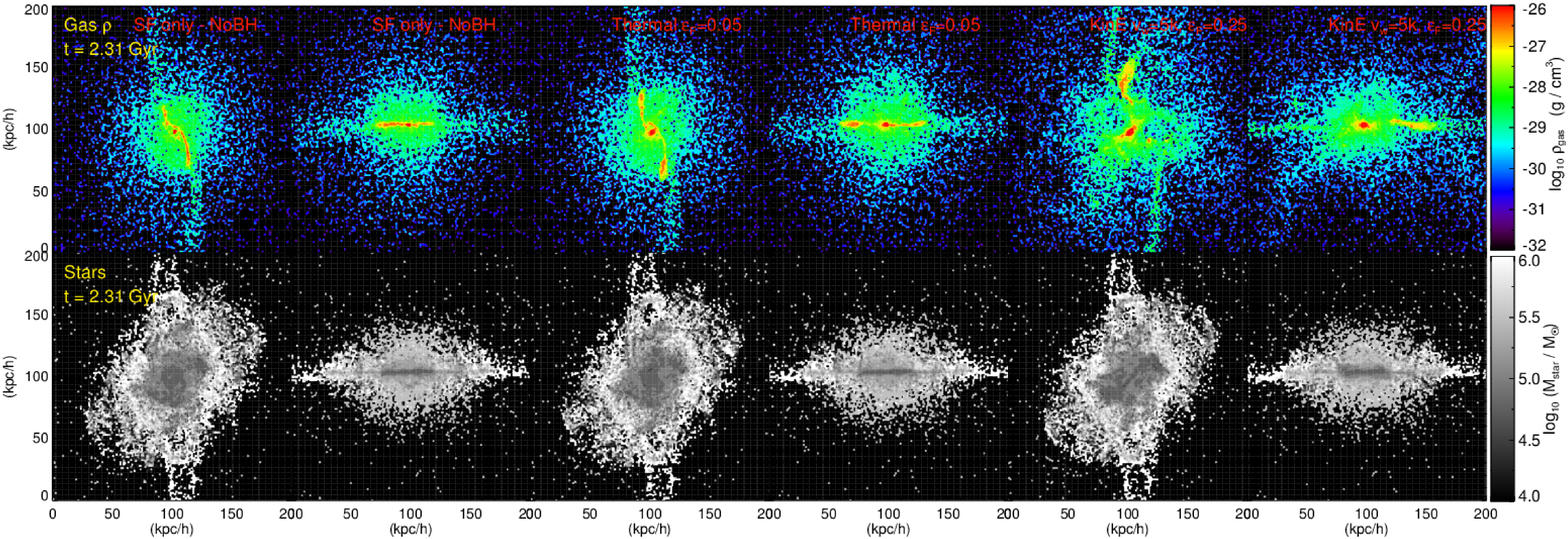} 
\end{array} 
$ 
\caption{ 
Projection of gas distribution and stars in fiducial galaxy merger simulation 
with different feedback models: 
{\it SF} (left two columns) - star formation only, 
{\it th3} (middle two columns) - BH thermal feedback with $\epsilon_f = 0.05$, 
and {\it kinE3} (right two columns) - BH kinetic feedback EDW with $v_w = 5000$ km/s and $\epsilon_f = 0.25$. 
The panels show the face-on (left) and edge-on (right) planes 
of a $(100 h^{-1}$ kpc$)^3$ volume at time $1.59$ Gyr in the top two rows, and 
a $(200 h^{-1}$ kpc$)^3$ volume at $t = 2.31$ Gyr in the bottom two rows. 
Projected gas density is plotted in the first and third rows, and 
projected total stellar mass in the second and fourth rows. 
} 
\label{fig-XY-YZ-Merger-snap57-snap83} 
\end{figure*} 

%%%%%%%%%%%%%%%%%%%%%%%%%%%%%%%%%%%%%%%%%%%%%%%%%%%%%%%%%%%%%%%%%%%%%%%% 

% The panels of Fig.~\ref{fig-XY-YZ-Merger-snap57-snap83} display the face-on and edge-on planes 
% of a $(100 h^{-1}$ kpc$)^3$ volume at time $1.59$ Gyr in the top two rows, and 
% a $(200 h^{-1}$ kpc$)^3$ volume at $t = 2.31$ Gyr in the bottom two rows. 
% Projected gas density is shown in the first and third rows, and 
% projected total stellar mass counting all stars 
% (disk, bulge, newly formed in the simulation from gas particles by active SF) in the second and fourth rows. 
% which can be seen as the gas off the disk plane and in the region between the two galaxies. 

Fig.~\ref{fig-XY-YZ-Merger-snap57-snap83} presents the projected distributions of gas and stars 
in the fiducial galaxy merger with three representative models: 
{\it SF} (left two columns) - star formation only, 
{\it th3} (middle two columns) - BH thermal feedback with $\epsilon_f = 0.05$, and 
{\it kinE3} (right two columns) - BH kinetic feedback EDW with $v_w = 5000$ km/s and $\epsilon_f = 0.25$. 
In the latter two runs the BHs grow to a comparable mass 
$M_{\rm BH} \sim 7 \times 10^6 M_{\odot}$ (from Fig.~\ref{fig-MassBH-SigmaStarLOS}). 
The intermittent gas outbursts produced by BH kinetic feedback, 
which were clearly distinguishable in the isolated galaxy, is difficult to disentangle in a merger, 
because other dynamical processes related to the merger process cause substantial gas to outflow. 

The top two rows ($t = 1.59$ Gyr) of Fig.~\ref{fig-XY-YZ-Merger-snap57-snap83} 
depict an epoch when the galaxies are approaching each other, 
on the way to their second pericenter passage and subsequent coalescence. 
Earlier at $t \sim 0.4$ Gyr the merging galaxy pair goes through a first pericenter passage 
during which there is a grazing collision of the outer disks. 
The resulting shocks and tidal interactions cause some gas to leave the disk plane and outflow, 
in all the models including the {\it SF} case. 
BH feedback induces enhanced gas outflow: 
more in run {\it th3} than {\it SF}, and highest in run {\it kinE3}. 
Each merging galaxy in all the runs exhibit spiral patterns composed of overdense gas. 
The spiral patterns are somewhat disturbed with kinetic BH feedback. 
The stellar distribution is almost indistinguishable between the feedback models here at $1.59$ Gyr. 

The bottom two rows ($t = 2.31$ Gyr) of Fig.~\ref{fig-XY-YZ-Merger-snap57-snap83} shows the 
resulting merged galaxy at an epoch during coalescence. 
It consist of a central compact spheroid and two tidal tails of overdense gas 
(visible as red in the third row, tails more prominent in {\it SF} and {\it th3} runs). 
There is a diffuse gaseous halo larger in size, which is more spherically shaped in the {\it SF} case, 
and quite disturbed giving an irregular appearance with kinetic feedback. 
The stellar distribution display a few differences between the models at $2.31$ Gyr: 
central $30 h^{-1}$ kpc radius of the face-on panels exhibit 
more spherically shaped structure in the {\it SF} case, then {\it th3}, and 
more elliptically shaped in {\it kinE3}.

\subsection{Black Hole Accretion Rate, Feedback Power, \& Star Formation Rate} 
\label{sec-res-BHAR-SFR} 

%%%%%%%%%%%%%%%%%%%%%%%%%%%%%%%%%%%%%%%%%%%%%%%%%%%%%%%%%%%%%%%%%%%%%%%% 

% FIGURE 4 
\begin{figure*} 
\centering 
\includegraphics[width = 1.0 \linewidth]{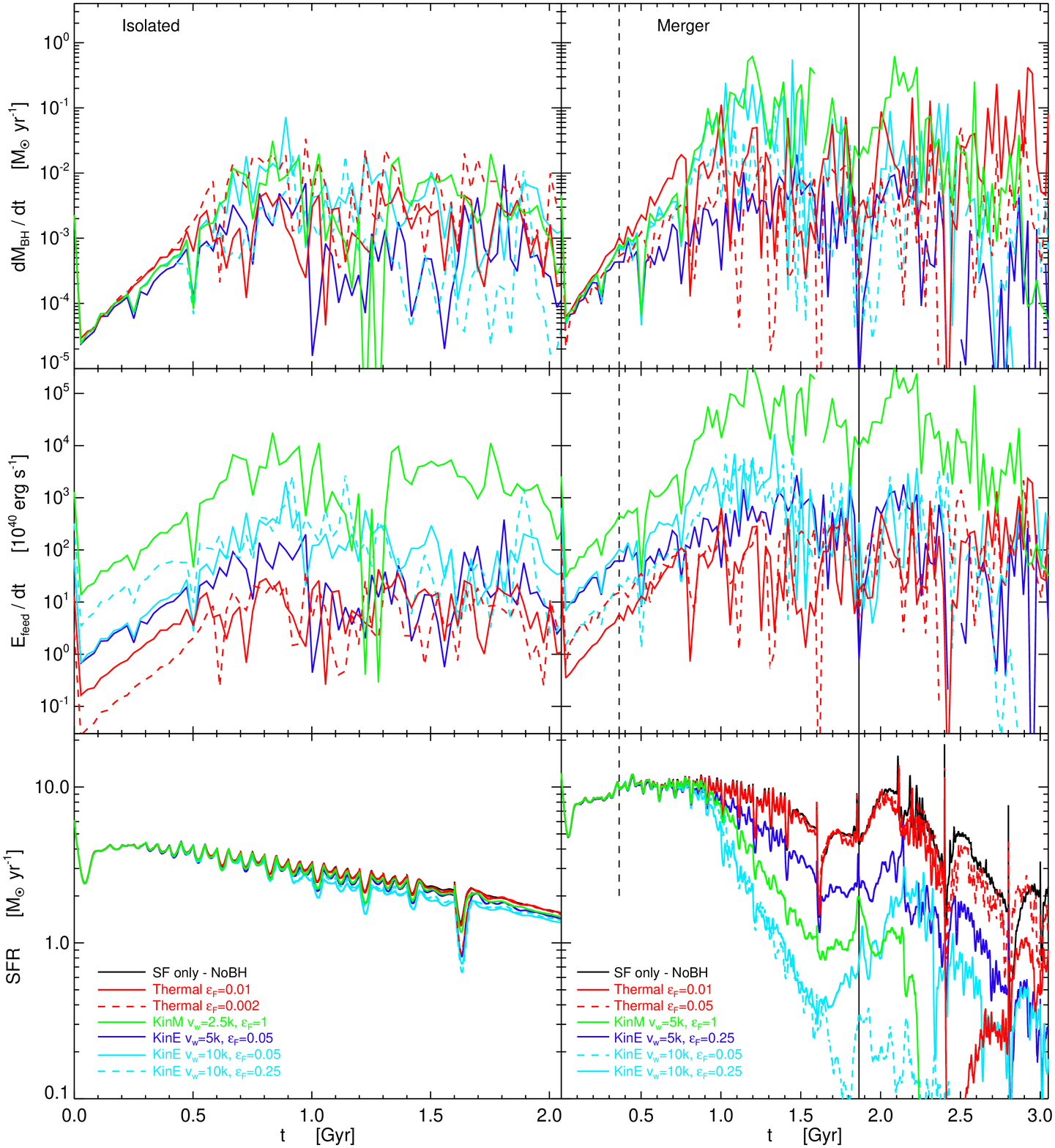} 
\caption{ 
Evolution with time of BH mass accretion rate ({\it top row}), energy feedback rate ({\it middle row}), 
and total star formation rate ({\it bottom row}), 
in the fiducial case of isolated galaxy ({\it left column}), and merger ({\it right column}) simulations. 
The different colours and plotting linestyles discriminate AGN feedback models labelled in the bottom panels. 
The runs for a single isolated galaxy are: 
{\it SF}    ({\it black solid}) - star formation only, 
{\it th1}   ({\it red dashed})  - thermal with $\epsilon_f = 0.002$, 
{\it th2}   ({\it red solid})   - thermal with $\epsilon_f = 0.01$, 
{\it kinE2} ({\it blue solid})  - kinetic EDW with $v_w =  5000$ km/s and $\epsilon_f = 0.05$, 
{\it kinE5} ({\it cyan solid})  - kinetic EDW with $v_w = 10000$ km/s and $\epsilon_f = 0.05$, 
{\it kinE6} ({\it cyan dashed}) - kinetic EDW with $v_w = 10000$ km/s and $\epsilon_f = 0.25$, 
{\it kinM3} ({\it green solid}) - kinetic MDW with $v_w =  2500$ km/s and $\epsilon_f = 1$. 
The contributions from each of the two merging galaxies are summed over, 
and the resulting total rates are plotted in the right column. 
Note that the parameter values of the plotted linestyles are different between isolated and merger cases. 
The galaxy merger runs are: 
{\it SF}    ({\it black solid}) - star formation only, 
{\it th2}   ({\it red solid})   - thermal with $\epsilon_f = 0.01$, 
{\it th3}   ({\it red dashed})  - thermal with $\epsilon_f = 0.05$, 
{\it kinE3} ({\it blue solid})  - kinetic EDW with $v_w =  5000$ km/s and $\epsilon_f = 0.25$, 
{\it kinE5} ({\it cyan dashed}) - kinetic EDW with $v_w = 10000$ km/s and $\epsilon_f = 0.05$, 
{\it kinE6} ({\it cyan solid})  - kinetic EDW with $v_w = 10000$ km/s and $\epsilon_f = 0.25$, 
{\it kinM2} ({\it green solid}) - kinetic MDW with $v_w =  5000$ km/s and $\epsilon_f = 1$. 
The vertical dashed black line in the right panels marks the epoch of 
first pericenter passage of the two merging galaxies, 
and the vertical solid black line marks the second pericenter passage. 
} 
\label{fig-BHAccr-StarForm-Rate} 
\end{figure*} 

% {\it kinE1} ({\it blue solid})   - kinetic EDW with $v_w =  5000$ km/s and $\epsilon_f = 0.01$, 
% {\it kinM2} ({\it green dashed}) - kinetic MDW with $v_w =  5000$ km/s and $\epsilon_f = 1$, 

% {\it kinE2} ({\it blue solid})   - kinetic EDW with $v_w =  5000$ km/s and $\epsilon_f = 0.05$, 
% {\it kinM3} ({\it green solid})  - kinetic MDW with $v_w =  2500$ km/s and $\epsilon_f = 1$. 

%%%%%%%%%%%%%%%%%%%%%%%%%%%%%%%%%%%%%%%%%%%%%%%%%%%%%%%%%%%%%%%%%%%%%%%% 

% labelled by the colours in the bottom-left panel. 
% On average the BHAR remains at almost steady state after $(0.7 - 0.8)$ Gyr up to $3$ Gyr. 
% Kinetic feedback produces a few times larger BHAR than thermal. 
% The maximum BHAR of the three runs are: 
% $0.005 M_{\odot}$/yr in {\it th2} ({\it red solid} curve), 
% $0.02 M_{\odot}$/yr in {\it kinM3} ({\it green solid}), and 
% $0.1 M_{\odot}$/yr in {\it kinE2} ({\it blue solid}). 
% depending on the feedback model; 
% with the AGN feedback models indicated by the different colours and plotting linestyles as labelled. 
% to $2.5 M_{\odot}$/yr at $0.05$ Gyr % to $4 M_{\odot}$/yr at $0.15$ Gyr. 
% remains constant at $4 M_{\odot}$/yr on average between $(0.15 - 0.5)$ Gyr, 

The BH mass accretion rate is an important quantity in the models, which governs BH growth 
as well as provides feedback energy in a self-regulated manner. 
It is measurable in observations, which can be compared to simulation results. 
The time evolution of BHAR in our fiducial isolated galaxy 
is presented in Fig.~\ref{fig-BHAccr-StarForm-Rate}, {\it top-left} panel. 
At $t = 0$, we find an accretion surge of $0.002 M_{\odot}$/yr 
because of our simulation initial condition. 
Embedding a $10^5 M_{\odot}$ BH in the gas-rich environment of a disk galaxy center suddenly, 
results in a high accretion rate. 
It reduces by $100$ times soon after, because of reduced central gas, 
which has depleted by the initial burst of accretion and SF. 
From $0.02$ Gyr, the BHAR rises linearly up to a time $(0.5 - 0.7)$ Gyr 
to reach a few times $0.001 - 0.01 M_{\odot}$/yr, 
the duration coinciding with the exponential mass growth of the BH 
(\S\ref{sec-res-Mass-Comp}, Fig.~\ref{fig-Evol-Gas-Star-BH-Isolated} top-left panel). 
There are heavy fluctuations in the BHAR, 
whereby it increases or decreases by a factor of up to $100$ in $0.02$ Gyr. 
The BHAR in terms of the Eddington mass accretion rate (Eq.~\ref{eq-LEdd}) 
also displays significant variability, especially at $t > 0.5$ Gyr. 
$\dot{M}_{\rm BH} / \dot{M}_{Edd}$ varies between $10^{-3} - 1$ for the thermal feedback models, 
while between $10^{-5} - 1$ for kinetic. 

The feedback energy rate ($\dot{E}_{\rm feed}$, Eq.~\ref{eq-Edot-Feed}) 
is plotted in Fig.~\ref{fig-BHAccr-StarForm-Rate}, {\it middle-left} panel. 
As expected the feedback power has same trends as BHAR, 
however the absolute value of $\dot{E}_{\rm feed}$ is different depending on $\epsilon_f$. 
At $t < 0.5$ Gyr, when the BH mass is small, higher $\epsilon_f$ produces a larger power. 
For the same feedback model, 
the impact of varying $\epsilon_f$ reduces at $t > 0.5$ Gyr, 
as the BHs grow, and the BHAR becomes a dominating factor in $\dot{E}_{\rm feed}$. 

The total star formation rate in the fiducial isolated galaxy versus time 
is displayed in the {\it bottom-left} panel of Fig.~\ref{fig-BHAccr-StarForm-Rate}. 
Similar to BHAR, at $t = 0$ there is an initial burst of SFR of $6 M_{\odot}$/yr. 
It reduces afterward via depletion of central gas, and then increases again. 
The SFR decreases linearly after $0.5$ Gyr, as more gas in the galaxy is converted to stars, 
to reach $1.5 M_{\odot}$/yr at $2$ Gyr. 
There are periodic fluctuations in the SFR, when it would decrease by a factor of $(1.1 - 2)$, 
occurring because of supernovae feedback in the stellar evolution model. 
All the AGN models show these general trends similarly. 
Thermal feedback produces almost the same SFR as the {\it SF} run. 
Kinetic feedback, more prominently the EDW models, 
produce up to $1.5$ times lower SFR than the {\it SF} and thermal cases at $t > 0.9$ Gyr. 

The {\it top-right} panel of Fig.~\ref{fig-BHAccr-StarForm-Rate} depicts the 
BH mass accretion rate time evolution in the fiducial galaxy merger, 
with the rates summed over each of the two merging BHs initially when they are separate. 
Note that the parameter values of the plotted linestyles are different between isolated and merger cases. 
Most of features of the BHAR in the merger are same as that of the isolated. 
The amplitude of fluctuations (change by a factor of up to $1000$ in $0.02$ Gyr) 
are larger in a merger than isolated case, 
because of extra dynamical processes (tidal forces, shocks) acting on the two merging galaxies. 
The vertical dashed and solid black lines in the right panels of Fig.~\ref{fig-BHAccr-StarForm-Rate} 
mark the epochs of first ($t \sim 0.36$ Gyr) and second ($t \sim 1.86$ Gyr) 
pericenter passages of the two galaxies, 
which occur at a concurrent time for the different feedback models. 
The galaxies merge at the second pericenter epoch, 
while the BHs undergo a third and some subsequent pericenters before merging. 

The feedback power in the merger is in Fig.~\ref{fig-BHAccr-StarForm-Rate}, 
{\it middle-right} panel. 
Its dependence on BHAR and $\epsilon_f$ are similar to that in an isolated galaxy. 

% The AGN feedback models are distinguished by the colours and linestyles, as labelled in the bottom-right panel. 
% the initial burst of accretion, sudden decrease, 
% linear increase between $\sim (0.02 - 1)$ Gyr coinciding with the exponential mass growth of the merging BHs 
% (\S\ref{sec-res-Mass-Comp}, Fig.~\ref{fig-Evol-Gas-Star-BH-Merger} top-left panel). 
% The merger gas dynamics also cause the BHAR to never reach any proper steady state, 
% contrary to the isolated galaxy. 

% The BHAR has 2 or 3 local peaks at different times,   
% revealing the varying amounts of gas inflow to the inner galaxy regions during the merger process. 
% The 1st peak represents the end of the exponential BH growth: 
% $0.1 M_{\odot}$/yr for run {\it th2} ({\it red solid} curve) at $1$ Gyr, 
% $1 M_{\odot}$/yr for run {\it kinM3} ({\it green solid}) 
% and $0.15 M_{\odot}$/yr for {\it kinE2} ({\it blue solid}) at $1.5$ Gyr. 
% The 2nd peak at $\sim (2 - 2.1)$ Gyr marks a BHAR of 
% $0.1 M_{\odot}$/yr in runs {\it th2} and {\it kinE2}, $2 M_{\odot}$/yr in {\it kinM3}; 
% this corresponds to the second pericenter passage of the galaxies, 
% when the nuclear regions actually collide causing substantial gas inflow to the center. 
% The 3rd peak occurs at $t > 2.5$ Gyr, after the 2 BHs have merged, 
% and indicate another episode of gas infall in the merged galaxy, 
% where the BHAR reaches $(0.1 - 0.4) M_{\odot}$/yr. 

The {\it bottom-right} panel of Fig.~\ref{fig-BHAccr-StarForm-Rate} shows 
the total star formation rate in the fiducial merger versus time, 
with the rates summed over each of the two galaxies. 
The initial features of the SFR in the merger case are the same as that of the isolated. 
The stellar evolution induced fluctuations have larger amplitudes 
at $t > 1$ Gyr in a merger than an isolated galaxy. 
The SFR decreases linearly from $0.9$ Gyr to $(1.6 - 1.7)$ Gyr reaching a 
local minimum depending on the AGN model. 
It rises subsequently because of additional gas inflow to dense central regions of the merging galaxies, 
and passes through a local peak at $t = 1.86$ Gyr during the second pericenter passage. 
It reaches another peak at $(2.1 - 2.2)$ Gyr, and decreases henceforth by gas depletion. 
At $t > 2$ Gyr, thermal feedback (red curves) 
produces lower SFR than the {\it SF} run (black curve) by a factor between $(1.5 - 10)$. 
Kinetic feedback causes a greater suppression of SFR, starting from $0.9$ Gyr; 
the reduction factors with respect to the {\it SF} case are: 
$5 - 30$ times in EDW models with $v_w = 5000$ km/s (blue) and in MDW models (green), 
$10 - 100$ times for EDW models with $v_w = 10000$ km/s (cyan). 

% the burst of SF at $t = 0$, and then sudden decrease. 
% There are 2  periods of SFR rise with different temporal slopes: 
% from $5 M_{\odot}$/yr at $0.05$ Gyr quickly to $8 M_{\odot}$/yr at $0.1$ Gyr, 
% and then has a slow growth to $10 M_{\odot}$/yr at $0.5$ Gyr. 
% The SFR increases afterward, and remains constant at $10 M_{\odot}$/yr on average between $(0.5 - 0.9)$ Gyr. 
% All the AGN models show these general trends similarly, and start to deviate from $0.9$ Gyr onwards. 
% when the nuclear regions collide, 

% $2 M_{\odot}$/yr for {\it SF} and thermal feedback, 
% $1 M_{\odot}$/yr for kinetic EDW with $v_w =  5000$ km/s and MDW, 
% $0.1 M_{\odot}$/yr for kinetic EDW with $v_w =  10000$ km/s. 

% $12 M_{\odot}$/yr for {\it SF} and thermal feedback, 
% $4 M_{\odot}$/yr for kinetic EDW with $v_w =  5000$ km/s and MDW. 
% This epoch corresponds to the second pericenter passage of the galaxies, 
% and the BHAR also has its 2nd peak. 

\subsection{Galaxy Mass Components} 
\label{sec-res-Mass-Comp} 

%%%%%%%%%%%%%%%%%%%%%%%%%%%%%%%%%%%%%%%%%%%%%%%%%%%%%%%%%%%%%%%%%%%%%%%% 

% FIGURE 5 
\begin{figure*} 
\centering 
\includegraphics[width = 1.0 \linewidth]{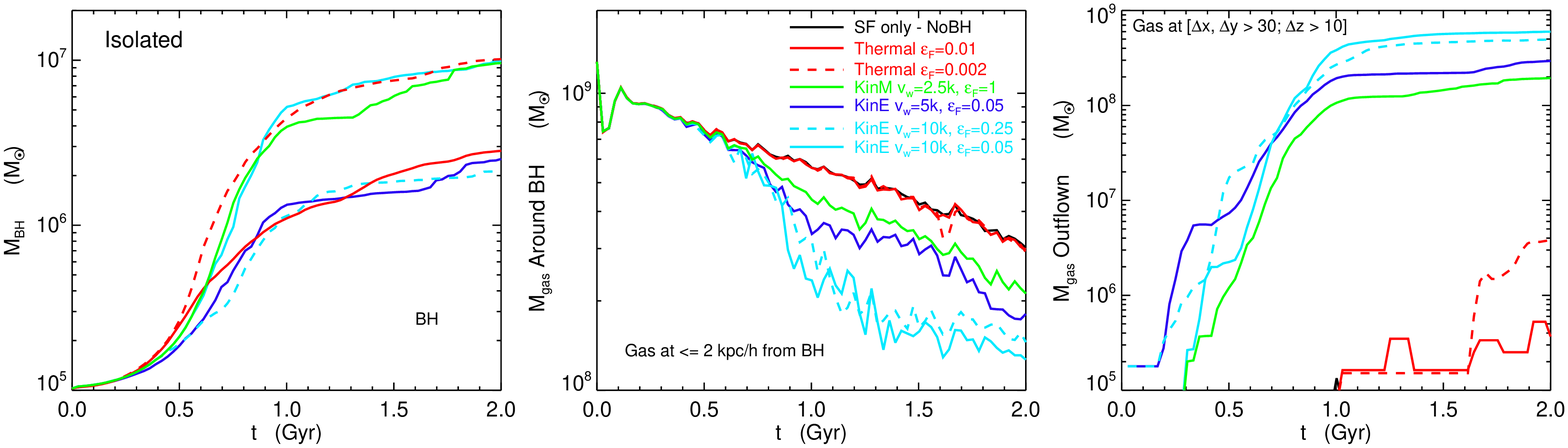} 
\caption{ 
Time evolution of component masses in isolated galaxy simulation, fiducial model: 
BH mass ({\it left} panel), 
gas mass inside distance $r \leq 2 h^{-1}$ kpc from BH ({\it middle}), 
gas mass outflown or that outside $\Delta x, \Delta y > 30 h^{-1}$ kpc 
and $\Delta z > 10 h^{-1}$ kpc ({\it right}). 
The distinguishing colours and plotting linestyles indicate different AGN feedback models: 
{\it SF}    ({\it black solid})  - star formation only, 
{\it th1}   ({\it red dashed})   - BH thermal feedback with $\epsilon_f = 0.002$, 
{\it th2}   ({\it red solid})    - BH thermal feedback with $\epsilon_f = 0.01$, 
{\it kinE2} ({\it blue solid})   - BH kinetic feedback EDW with $v_w =  5000$ km/s and $\epsilon_f = 0.05$, 
{\it kinE5} ({\it cyan solid})   - BH kinetic feedback EDW with $v_w = 10000$ km/s and $\epsilon_f = 0.05$, 
{\it kinE6} ({\it cyan dashed})  - BH kinetic feedback EDW with $v_w = 10000$ km/s and $\epsilon_f = 0.25$, 
{\it kinM3} ({\it green solid})  - BH kinetic feedback MDW with $v_w =  2500$ km/s and $\epsilon_f = 1.0$. 
} 
\label{fig-Evol-Gas-Star-BH-Isolated} 
\end{figure*} 

%mass of new stars formed from gas via SF ({\it bottom-left}), 
%total gas mass in galaxy ({\it bottom-middle}), 
%gas mass inside disk or that at $\Delta x, \Delta y \leq 30 h^{-1}$ kpc 
%and $\Delta z \leq 10 h^{-1}$ kpc ({\it bottom-right}). 
%{\it kinE1} ({\it blue solid})   - BH kinetic feedback EDW with $v_w =  5000$ km/s, $\epsilon_f = 0.01$, 
%{\it kinM2} ({\it green dashed}) - BH kinetic feedback MDW with $v_w =  5000$ km/s, $\epsilon_f = 1.0$, 

%%%%%%%%%%%%%%%%%%%%%%%%%%%%%%%%%%%%%%%%%%%%%%%%%%%%%%%%%%%%%%%%%%%%%%%% 

The masses of BH and gas components in the isolated fiducial galaxy models 
versus evolution time is plotted in Fig.~\ref{fig-Evol-Gas-Star-BH-Isolated}. 
The {\it left} panel presents the BH mass, 
where all the feedback models have the BH growing in a qualitatively similar manner. 
Starting from a seed mass of $10^5 M_{\odot}$, each BH first undergoes a slow growth. 
It then has an exponential growth over the time range $(0.5 - 1)$ Gyr, 
when its mass increases by a factor $10 -$ a few $100$. 
After $1$ Gyr it comes to an almost steady state, having a very slow subsequent growth. 
The final BH mass reached at $2$ Gyr depends on the AGN model, 
and is inversely proportional to $\epsilon_f$ and directly proportional to $v_w$. 
A higher $\epsilon_f$ imparts a stronger feedback affecting more central gas, 
and yields a less-massive BH than a lower $\epsilon_f$. 
On increasing $v_w$, $\dot{M}_w$ decreases 
(inversely proportionality in Eq.~(\ref{eq-MdotW-EDW}) for EDW, and (\ref{eq-MdotW-MDW}) for MDW), 
there is reduced kinetic feedback and less gas is ejected out, 
making more gas available for accreting onto BH which grows more massive. 
As a note, a BH which is only accreting gas with no feedback 
grows to a mass of $3 \times 10^8 M_{\odot}$ in $2$ Gyr.

The {\it right} panel of Fig.~\ref{fig-Evol-Gas-Star-BH-Isolated} 
depicts the gas mass which has outflown, or that lying outside a pre-defined disk region. 
No gas outflows in the {\it SF} run. 
In the case of thermal feedback a tiny fraction ($< 10^{-3}$) of gas outflows after $1$ Gyr. 
The kinetic feedback models cause some gas to outflow starting from $(0.2 - 0.3)$ Gyr. 
The outflowing mass rises exponentially during the peak period of BH growth, 
because the feedback energy governing the mass outflow is derived from the BH accretion rate. 
A few $\times 10^8 M_{\odot}$ gas has outflown by $1$ Gyr, 
which is a fraction $0.03 - 0.07$ of the initial gas mass. 

The {\it middle} panel of Fig.~\ref{fig-Evol-Gas-Star-BH-Isolated} denotes the time evolution of central gas. 
This summed mass inside $r \leq 2 h^{-1}$ kpc from the BH 
consist of the innermost gas which is accreted onto the BH, ejected out by kinetic feedback, 
and is part of the central high-density star-forming region of the galaxy where gas is converted to stars. 
This mass decreases equally for all the feedback models up to $\sim 0.7$ Gyr, 
which is an impact of star formation, and the BH mass is still small. 
Thermal feedback does not alter the central gas reservoir content. 
In the kinetic models, the inner gas is depleted such that the central mass becomes 
$0.25 - 0.5$ of the {\it SF} case between $(1 - 2)$ Gyr. 
There the BH grows appreciably after $0.5$ Gyr, exerts strong kinetic feedback, and ejects the central gas out.

%%%%%%%%%%%%%%%%%%%%%%%%%%%%%%%%%%%%%%%%%%%%%%%%%%%%%%%%%%%%%%%%%%%%%%%% 

% FIGURE 6 
\begin{figure*} 
\centering 
\includegraphics[width = 1.0 \linewidth]{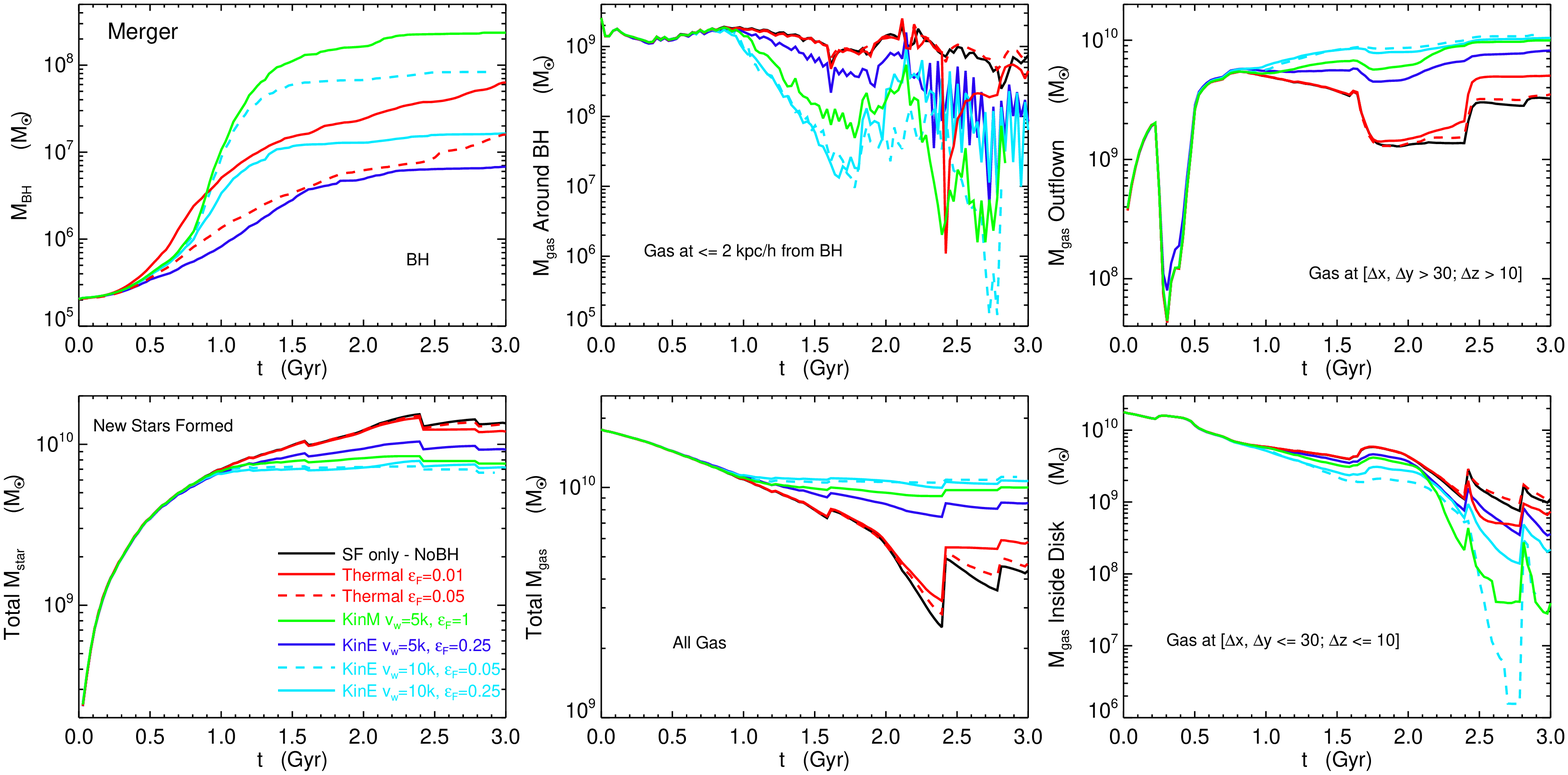} 
\caption{ 
Time evolution of various component (BH, gas, stars)'s masses in galaxy merger simulation, fiducial model: 
BH mass ({\it top-left} panel), 
gas mass inside $r \leq 2 h^{-1}$ kpc from BH ({\it top-middle}), 
gas mass outflown ({\it top-right}), 
mass of new stars formed from gas ({\it bottom-left}), 
total gas mass in galaxy ({\it bottom-middle}), 
gas mass inside disk ({\it bottom-right}). 
The mass component for each of the two merging galaxies are summed over, 
and the resulting total masses are plotted. 
The different colours and plotting linestyles distinguish between AGN feedback models: 
{\it SF}    ({\it black solid}) - star formation only, 
{\it th2}   ({\it red solid})   - BH thermal feedback with $\epsilon_f = 0.01$, 
{\it th3}   ({\it red dashed})  - BH thermal feedback with $\epsilon_f = 0.05$, 
{\it kinE3} ({\it blue solid})  - BH kinetic feedback EDW with $v_w =  5000$ km/s and $\epsilon_f = 0.25$, 
{\it kinE5} ({\it cyan dashed}) - BH kinetic feedback EDW with $v_w = 10000$ km/s and $\epsilon_f = 0.05$, 
{\it kinE6} ({\it cyan solid})  - BH kinetic feedback EDW with $v_w = 10000$ km/s and $\epsilon_f = 0.25$, 
{\it kinM2} ({\it green solid}) - BH kinetic feedback MDW with $v_w =  5000$ km/s and $\epsilon_f = 1.0$. 
} 
\label{fig-Evol-Gas-Star-BH-Merger} 
\end{figure*} 

% in a similar format as Fig.~\ref{fig-Evol-Gas-Star-BH-Isolated} 
% {\it kinE2} ({\it blue solid})   - BH kinetic feedback EDW with $v_w =  5000$ km/s, $\epsilon_f = 0.05$, 
%{\it kinM3} ({\it green solid})  - BH kinetic feedback MDW with $v_w =  2500$ km/s $\epsilon_f = 1.0$. 

%%%%%%%%%%%%%%%%%%%%%%%%%%%%%%%%%%%%%%%%%%%%%%%%%%%%%%%%%%%%%%%%%%%%%%%% 

Fig.~\ref{fig-Evol-Gas-Star-BH-Merger} displays the time evolution of 
various component (BH, gas, stars)'s masses in the fiducial merger simulation. 
The mass contribution from each of the two merging galaxies are summed over, 
and the resulting total masses are plotted. 
Some trends are similar to the isolated galaxy evolution in Fig.~\ref{fig-Evol-Gas-Star-BH-Isolated}, 
while some are different caused by the dynamics of the merger process. 

The BH mass in the {\it top-left} panel of Fig.~\ref{fig-Evol-Gas-Star-BH-Merger} 
shows a similar general trend as isolated galaxy. 
The steady state reached at $t > 1.5$ Gyr, after the exponential growth, 
is more prominent with kinetic feedback than thermal. 
The impact of varying $v_w$ is reduced in a merger than a single isolated galaxy. 

% The galaxy undergoes a stronger feedback with a higher value of $\epsilon_f$, 
% and produces a less-massive BH than a lower $\epsilon_f$; 
% seen in the solid versus dashed red and cyan curves. 
% slow growth up to $t \sim 0.5$ Gyr, and exponential growth between $(0.5 - 1.5)$ Gyr. 
% The AGN feedback model is distinguished by the colour and plotting linestyle, labelled in the bottom-left panel. 
% With kinetic EDW $\epsilon_f = 0.25$ case, $v_w = 10000$ km/s (cyan solid) produces just a 
% $\sim 2$ times more massive BH than $v_w =  5000$ km/s (blue solid curve). 
% While the effect is opposite for kinetic MDW (green curves), 
% $v_w = 5000$ km/s (dashed) produces a BH of mass $0.7$ times than that 
% with $v_w =  2500$ km/s (solid) at $3$ Gyr. 
% at $\Delta x, \Delta y > 30 h^{-1}$ kpc and $\Delta z > 10 h^{-1}$ kpc. 
% , and returns to the original trend at $0.5$ Gyr. 
% increasing the disk mass and reducing the outflow mass. 
% It remains nearly constant for all the feedback models up to $1$ Gyr. 

The mass of new stars formed ({\it bottom-left} panel of Fig.~\ref{fig-Evol-Gas-Star-BH-Merger}) 
increases overall, 
and the total gas mass in the galaxies ({\it bottom-middle}) decreases 
with simulation time over $(0 - 3)$ Gyr, for all the AGN models, 
as gas is converted to stars and is accreted onto central BH. 
Except there are some epochs (at $\sim 1.2, 1.6, 2.4, 2.8$ Gyr) when the star mass reduces 
and gas mass increases by the same amount, occurring because of stellar mass loss 
(a part of the stellar evolution model of \citealt{Tornatore07}). 
Thermal feedback produces a late visible impact; 
at $t > 2$ Gyr there is slightly smaller star mass and a higher gas mass than the {\it SF} case. 
The influence of kinetic feedback is larger, visible from $1.1$ Gyr; 
the stellar mass is reduced by a factor of $1.7 - 2.5$ than {\it SF}, 
resulting in a correspondingly higher gas mass. 

The {\it bottom-right} panel of Fig.~\ref{fig-Evol-Gas-Star-BH-Merger} shows the gas mass inside a disk region, 
within distances from central BH $\Delta x, \Delta y \leq 30 h^{-1}$ kpc along the disk plane, 
and $\Delta z \leq 10 h^{-1}$ kpc in the perpendicular direction. 
At $t > 1$ Gyr, the kinetic feedback cases have a smaller disk mass 
than the {\it SF} and thermal runs, occurring because of increased outflows. 

The {\it top-right} panel of Fig.~\ref{fig-Evol-Gas-Star-BH-Merger} 
denotes the gas mass which has outflown, i.e. lying outside the disk region. 
Contrary to the isolated case, 
in the merger of two galaxies there is significant gas outflow from the beginning, for all the models. 
The {\it SF} run has $3 \times 10^9 M_{\odot}$ gas outflown by $3$ Gyr, 
a few times larger than the gas mass within the disk at the same time. 
This no-feedback outflow occurs mainly because of shock heating and collisions during the merger. 
In the case of thermal feedback 
the outflown gas mass is up to two times higher than the {\it SF} case at $t > 2$ Gyr. 
The kinetic feedback models expel a significantly higher fraction of gas 
than the {\it SF} and thermal runs; $\sim 10^{10} M_{\odot}$ gas outflows by $3$ Gyr, 
which is comparable or larger than the mass of new stars formed, 
and $(100 - 1000)$ times higher than the gas mass remaining in the disk. 
Thus the resulting galaxy, produced by the merger of two disk galaxies, 
contains most of its gas in the form of an extended spheroidal halo. 

The rearrangement of gas distribution in the two galaxies during the merger 
is indicated by a few opposite trends in the component evolution. 
The outflow mass ({\it top-right} panel of Fig.~\ref{fig-Evol-Gas-Star-BH-Merger}) 
decreases sharply at $0.3$ Gyr, coinciding with a rise in disk mass ({\it bottom-right}). 
This corresponds to the first pericenter passage 
(marked in the right panels of Fig.~\ref{fig-BHAccr-StarForm-Rate}) 
of the galaxies approaching each other. 
During this encounter, tidal forces cause more gas to flow inward. 
After $0.5$ Gyr the galaxies move apart, restoring the original gas configuration in each. 
They undergo a second pericenter passage starting at $\sim 1.6$ Gyr, 
when the outflow mass reduces and disk mass increases again. 
The galaxy disks collide during this passage, the disk structures are lost in the violent encounter, 
some gas recoil forming tidal tails, 
and the nuclear regions merge between $(2 - 2.5)$ Gyr, making a galaxy with spheroidal gas distribution. 

The {\it top-middle} panel of Fig.~\ref{fig-Evol-Gas-Star-BH-Merger} shows the central gas evolution, 
inside $r \leq 2 h^{-1}$ kpc from the BH in each galaxy. 
It shows heavy fluctuations at $t > 1$ Gyr; 
falling because of accreting onto BH and/or ejecting out by feedback and merger shock heating, 
while rising again when there is gas inflow to the inner regions. 
The thermal feedback and {\it SF} cases are almost coincident up to $2.4$ Gyr, 
after which a strong burst of feedback reduces the central gas mass 
and increases the outflow mass in the {\it th2} run. 
In the kinetic models, the inner gas is depleted more at $1.5$ Gyr, replenished again later, 
presenting oscillatory behaviour between $(1.5 - 3)$ Gyr; 
finally the central mass becomes $(0.1 - 0.01)$ of the {\it SF} case at $3$ Gyr. 

We see, from Figs.~\ref{fig-BHAccr-StarForm-Rate}, \ref{fig-Evol-Gas-Star-BH-Isolated} 
and \ref{fig-Evol-Gas-Star-BH-Merger}, 
that thermal feedback produces negligible impact compared to the {\it SF} case 
in the isolated galaxy and at $t < 2.4$ Gyr in the merger. 
We investigate and infer this is because the thermal energy injected is radiated away very quickly 
by the dense star-forming gas particles in the multiphase model \citep{SH03}.

\subsection{Radial-Profiles} 
\label{sec-res-Radial-Profiles} 

%%%%%%%%%%%%%%%%%%%%%%%%%%%%%%%%%%%%%%%%%%%%%%%%%%%%%%%%%%%%%%%%%%%%%%%% 

% FIGURE 7 
\begin{figure*} 
\centering 
\includegraphics[width = 1.0 \linewidth]{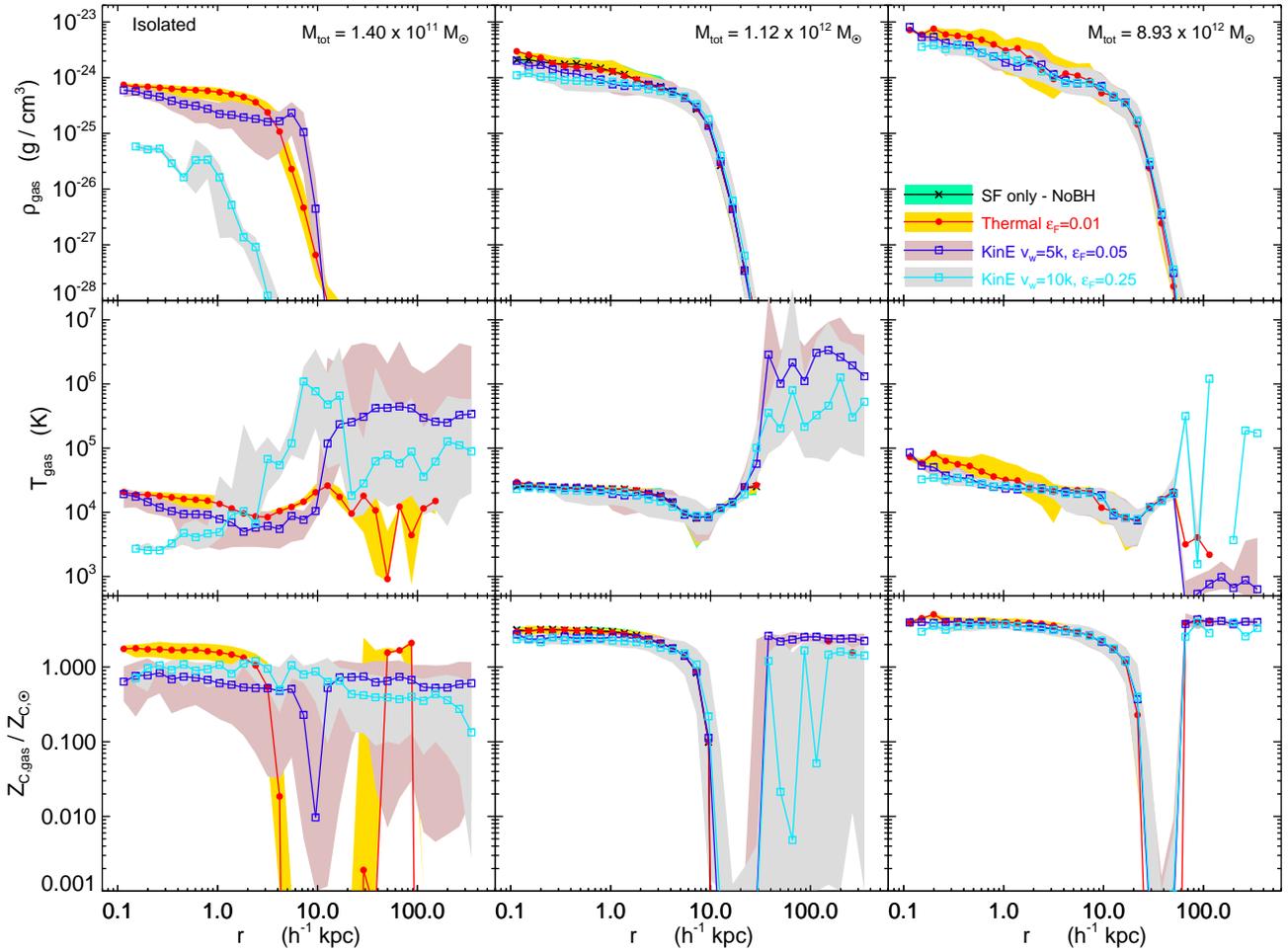} 
\caption{ 
Radial gas profiles of isolated galaxy models at an evolution time of $2$ Gyr.     % /snap_072/Gas/ 
The three columns indicate varying total galaxy mass: 
lower-mass with $v_{200} = 75$ km/s and $M_{\rm tot} = 1.40 \times 10^{11} M_{\odot}$ ({\it left}), 
fiducial with $v_{200} = 150$ km/s and $M_{\rm tot} = 1.12 \times 10^{12} M_{\odot}$ ({\it middle}), and 
higher-mass with $v_{200} = 300$ km/s and $M_{\rm tot} = 8.93 \times 10^{12} M_{\odot}$ ({\it right}). 
The AGN feedback models providing best-fit 
in the $[M_{\rm BH} - \sigma_{\star}]$ diagram of an isolated case are plotted: 
{\it SF}    ({\it black cross, green shade})        - star formation only, 
{\it th2}   ({\it red filled circle, yellow shade}) - thermal with $\epsilon_f = 0.01$, 
{\it kinE2} ({\it blue square, reddish grey shade}) - kinetic EDW with $v_w =  5000$ km/s and $\epsilon_f = 0.05$, 
{\it kinE6} ({\it cyan square, grey shade})         - kinetic EDW with $v_w = 10000$ km/s and $\epsilon_f = 0.25$. 
The plotted curves denote the median quantity in radial bins of each run. 
The respective shaded areas enclose the $70$th percentiles 
above and below the median in each case, showing the radial scatter. 
Top row is the gas density in cgs units. 
Middle row is gas temperature, and the effective equation-of-state $T$ is shown for the multiphase gas particles. 
Bottom row is carbon metallicity, showing the ratio of carbon mass fraction in gas to that of the Sun. 
} 
\label{fig-Radial-Profiles-Isolated} 
\end{figure*} 

%%%%%%%%%%%%%%%%%%%%%%%%%%%%%%%%%%%%%%%%%%%%%%%%%%%%%%%%%%%%%%%%%%%%%%%% 
%%%%%%%%%%%%%%%%%%%%%%%%%%%%%%%%%%%%%%%%%%%%%%%%%%%%%%%%%%%%%%%%%%%%%%%% 

% FIGURE 8 
\begin{figure*} 
\centering 
\includegraphics[width = 1.0 \linewidth]{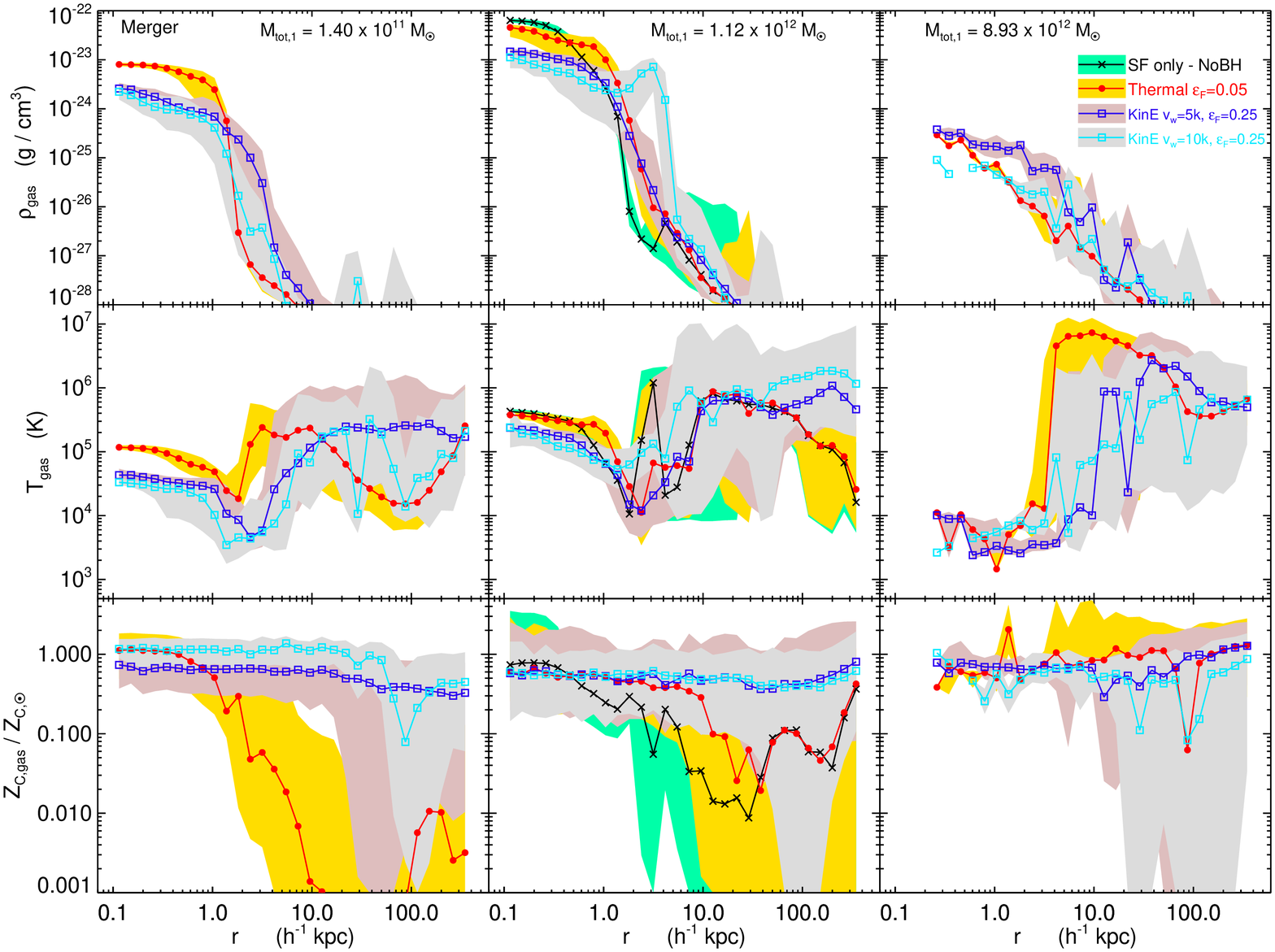} 
\caption{ 
Radial gas profiles of merged galaxy at an evolution time of $2.31$ Gyr.     % /snap_083/Gas/ 
The three columns indicate varying total mass of an individual galaxy undergoing merger. 
The AGN feedback models providing best-fit 
in the $[M_{\rm BH} - \sigma_{\star}]$ diagram of a merger are plotted: 
{\it SF}    ({\it black cross, green shade})        - star formation only, 
{\it th3}   ({\it red filled circle, yellow shade}) - thermal with $\epsilon_f = 0.05$, 
{\it kinE3} ({\it blue square, reddish grey shade}) - kinetic EDW with $v_w =  5000$ km/s and $\epsilon_f = 0.25$, 
{\it kinE6} ({\it cyan square, grey shade})         - kinetic EDW with $v_w = 10000$ km/s and $\epsilon_f = 0.25$. 
The plotted curves denote the median quantity in radial bins of each run, and 
the shaded areas enclose the $70$th percentiles above and below the median. 
Top row is gas density, 
middle row is gas temperature (with the effective $T$ shown for the multiphase gas particles), 
and bottom row is carbon metallicity. 
} 
\label{fig-Radial-Profiles-Merger} 
\end{figure*} 

% lower-mass with $v_{200} = 75$ km/s and $M_{\rm tot} = 1.40 \times 10^{11} M_{\odot}$ ({\it left}), 
% fiducial with $v_{200} = 150$ km/s and $M_{\rm tot} = 1.12 \times 10^{12} M_{\odot}$ ({\it middle}), and 
% higher-mass with $v_{200} = 300$ km/s and $M_{\rm tot} = 8.93 \times 10^{12} M_{\odot}$ ({\it right}). 

%%%%%%%%%%%%%%%%%%%%%%%%%%%%%%%%%%%%%%%%%%%%%%%%%%%%%%%%%%%%%%%%%%%%%%%% 

The radial profiles of gas properties of the isolated galaxy cases at an evolution time of $2$ Gyr 
are presented in Fig.~\ref{fig-Radial-Profiles-Isolated}. 
The radius is computed by the distance from the maximum gas density location. 
The total galaxy mass varies between the three columns: 
lower-mass of $M_{\rm tot} = 1.40 \times 10^{11} M_{\odot}$ ({\it left}), 
fiducial of $M_{\rm tot} = 1.12 \times 10^{12} M_{\odot}$ ({\it middle}), and 
higher-mass of $M_{\rm tot} = 8.93 \times 10^{12} M_{\odot}$ ({\it right}). 
Fig.~\ref{fig-Radial-Profiles-Merger} displays the radial gas profiles of the merged galaxy at $t = 2.31$ Gyr, 
an epoch when there is still significant gas left near the center, 
however the two BHs in a few runs have not merged yet. 
The total mass of an individual galaxy undergoing merger varies between the three columns. 
The profiles of the different models vary, more in the merged galaxy than in the isolated evolution. 
This is because of the distinct operation of each feedback, 
affecting the gas in diverse ways enhanced by a merger. 

% considered as the galaxy center. 
% The plotted solid curve denotes 
% the median value in radial bins for each run labelled by the colour and plotting symbol. 
% The grey shaded area encloses the $70$th percentiles 
% above and below the median in run {\it kinE3} (blue curve), showing the typical radial scatter. 
% which are broadly monotonic functions composed of different negative-sloped regions 
% separated by threshold radii dependent on galaxy mass and feedback model. 

\subsubsection{Density}
\label{sec-res-Radial-Density} 

The gas density radial profiles are plotted in the top rows. 
The isolated models in Fig.~\ref{fig-Radial-Profiles-Isolated} exhibit similar density profiles, 
except the lower-mass galaxy, 
where strong kinetic feedback in the {\it kinE6} run expels large amounts of gas and 
reduces the density by $10 - 100$ times than the other runs. 

In the merger, Fig.~\ref{fig-Radial-Profiles-Merger}, 
the lower-mass and fiducial galaxies present qualitatively similar profiles. 
However there are two exceptions in the fiducial galaxy: at $r \sim 3 h^{-1}$ kpc 
the {\it kinE6} case produces a local peak, while the {\it SF} run generates a local drop. 
In the inner $r < 2 h^{-1}$ kpc, 
the {\it SF} case exhibit a $\sim 1.5$ times higher density than that with thermal feedback, 
which in turn is $(3 - 5)$ times denser than those with kinetic feedback. 
The differences reduce and/or reverse in the outer $r > (2 - 3) h^{-1}$ kpc, 
where in the lower-mass galaxy kinetic feedback causes $\sim 10$ times more density than thermal. 
The higher-mass galaxy (top-right panel of Fig.~\ref{fig-Radial-Profiles-Merger}) 
has a smaller density than either of the two other galaxies. 
This implies that most of the gas has been expelled at $2.31$ Gyr here by strong feedback effects. 

The features are signatures of kinetic feedback being able to expel gas 
efficiently from central regions, consequently reducing the central gas density, 
and depositing the gas at larger distances. 

% The regions at $r < 0.8 h^{-1}$ kpc is denser in cases {\it kinE3} and {\it th3} than in {\it kinE6}. 
% While at $r > 1 h^{-1}$ kpc, models {\it kinE3} and {\it kinE6} cause higher density than {\it th3}. 

\subsubsection{Temperature} 
\label{sec-res-Radial-T} 

The temperature radial profiles are presented in the middle rows of 
Figs.~\ref{fig-Radial-Profiles-Isolated} and \ref{fig-Radial-Profiles-Merger}, 
using the effective temperature for those gas particles which are star-forming. 
The $T$-profiles in the inner regions $r \leq (1 - 2) h^{-1}$ kpc of all the galaxies 
follow the negative-sloped density-profiles (top rows in respective figures). 
This represents the dense gas at galaxy center undergoing SF, and 
having a $T$ between $\sim (10^{3} - 5 \times 10^{5})$ K, 
which is directly proportional to the density, as a result of following the 
SF effective equation of state in the multiphase model by \citet{SH03}. 

In the isolated (Fig.~\ref{fig-Radial-Profiles-Isolated}) kinetic feedback runs 
(except case {\it kinE2} in the higher-mass galaxy), 
the gas $T$ increases with radius at $r > (5 - 10) h^{-1}$ kpc, 
as the high-velocity outflows thermalize their energies at large-radii. 
The $T$ profiles reveal that thermal BH feedback does not 
heat up the inner gas to higher temperatures as a physical model should do. 

The fiducial merged galaxy (Fig.~\ref{fig-Radial-Profiles-Merger}) {\it SF} case 
has a smaller (than the feedback runs) cold central spheroidal core, 
surrounded by a hot gas halo of size $r \sim 3 h^{-1}$ kpc; 
and a prominent a cold annular ring at $r \sim (4 - 6) h^{-1}$ kpc 
composed of gas infalling from two tidal tails. 
The gas $T$ increases with radius at $r \geq 2 h^{-1}$ kpc. 
These outer regions contain gas shock heated during the merger, and 
feedback processed material: gas heated by BH thermal coupling, and shock heating via kinetically kicked gas. 
The $T$ profiles show a local peak, which for the 
{\it SF} and thermal feedback occurs at a radius (e.g., $r \sim 15 h^{-1}$ kpc in the fiducial galaxy) 
$10$ times smaller than the kinetic feedback models ($r \sim 200 h^{-1}$ kpc). 
This is because the jet-like outflows of kinetic feedback deposit their energies at a larger radii. 

% There is a change in $T$ slope in the outer regions, 

\subsubsection{Carbon Metallicity} 
\label{sec-res-Radial-ZC} 

%The dashed curves represent median-$Z_C$ for the enriched particles only, 
%i.e. those having $Z_C > 0$, without counting particles with $Z_C = 0$. 
% the profiles show decreasing $Z_C$ going outward from center at $r < 40 h^{-1}$ kpc 
% with varying $r$-dependent negative slopes, 
% and the $Z_C$ profiles start to rise again from $r > 40 h^{-1}$ kpc. 
% except a $(2 - 3)$ times higher $Z_C$ between $r \sim (10 - 70) h^{-1}$ kpc. 
% negative sloped profile in the lower-mass galaxy (left panel), 
% and positive slope in the fiducial and higher-mass galaxies (middle and right). 

Radial profiles of the ratio of carbon mass fraction in the gas to that of the Sun are plotted in the 
bottom rows of Figs.~\ref{fig-Radial-Profiles-Isolated} and \ref{fig-Radial-Profiles-Merger}. 
The medians and percentiles are computed considering all 
(both enriched and non-enriched) gas particles in radial bins. 
In the isolated galaxy (Fig.~\ref{fig-Radial-Profiles-Isolated}) the $Z_C$ profiles are flat overall. 
It has a sharp decrease at a mass-dependent radii corresponding to the end of the galaxy disk, 
and increases again at larger radii in the feedback runs comprising of metal-enriched gas ejected out. 

In the merged galaxy (Fig.~\ref{fig-Radial-Profiles-Merger}), 
the kinetic feedback models produce an 
almost flat, no-gradient $Z_C$ profile in the inner $r < 20 h^{-1}$ kpc; 
and small gradients at $r > 20 h^{-1}$ kpc. 
These features demonstrate that kinetic feedback 
is substantially effective in transporting metals away from central SF regions, 
and spreading them in the lower-density surrounding CGM. 
The {\it SF} and thermal feedback cases exhibit metallicity gradients at all-$r$ 
in the lower-mass and fiducial galaxies. 
Thermal feedback in the higher-mass galaxy produces a no-gradient flat $Z_C$ profile 
like the kinetic models, with comparable metallicity values. 
This implies thermal feedback is relatively more efficient to enrich the CGM of higher mass galaxies. 

In the inner $r < 0.5 h^{-1}$ kpc 
all the feedback models generate a $1.5$ times lower $Z_C$ than the {\it SF} case, 
because of suppression of central SF. 
The trend reverses in the outer $r > 0.5 h^{-1}$ kpc: 
thermal feedback displays a higher $Z_C$ up to $10$ times more than {\it SF}, 
and the kinetic cases exhibit up to $60$ times higher $Z_C$, 
originating from the accumulation of metal-enriched gas expelled by feedback.

\section{Discussion} 
\label{sec-discussion} 

We see that BH kinetic feedback operates intermittently, 
driving bipolar outflows visible for short times. 
The geometry of the outflowing gas varies with time and feedback model: 
from jet-like, perfectly collimated fast gas outflows, to wide-angled relatively-slower outflows. 
Sometimes precession of the jets are visible. 
The variety of outflow geometry can be seen as movies in the given 
weblink\footnote{http://adlibitum.oats.inaf.it/barai/AllPages/Visualization/Isolated\_Galaxy\_AGN.html}. 
The jet-like collimated outflows efficiently pierce through the galaxy ISM 
and escape out perpendicular to the disk. 
The jets are not perturbed by the environment, since there is no ambient gas 
outside the galaxy disk in our isolated cases. 
Effects like mass entrainment, bending and fragmenting of the AGN jets become important 
in different environments, e.g. for central galaxies in clusters. 

The rightmost column of Fig.~\ref{fig-XY-YZ-Isolated} represents a typical jet-like outburst example. 
% the evolution time of $t = 1.73$ Gyr was chosen for this figure to coincide with this. 
When gas accumulates near the galaxy center, the BHAR rises 
and feedback acts to eject gas out of the disk, 
consisting of multi-temperature, metal-enriched outflow, moving at a high speed $v_w$. 
This depletes (partially) the central gas reservoir, and the BHAR reduces, stopping the bipolar outflow. 
During the off-period, which occurs for a longer time than the outflow was on, 
gas flows in near the BH again. 
The process acts periodically with intermittent active outbursts, 
separated by longer quiescent intervals. 
% as long as the reservoir of gas in the central disk region is not depleted significantly. 
Whereas for BH thermal feedback, the outflow occurs relatively more continuously, 
and after a while some of the outflowing gas slows down, reverses and inflows. 

In cosmological simulations,  % of kpc-scale resolution, 
wind particles are often ejected perpendicular to the galaxy disk to generate bipolar outflows, 
e.g., that done for SNe-driven kinetic feedback \citep{Tescari09, Barai13}. 
We follow the same for AGN kinetic feedback here (\S\ref{sec-num-Implement}). 
As a test we perform a new run where the wind particles are ejected isotropically. 
% along a random direction around the BH. 
We find that in isolated galaxies (higher resolution here than in typical cosmological simulations), 
the effectiveness of kinetic feedback is almost equivalent for bipolar versus isotropic wind ejection. 
The BHAR, SFR, stellar mass and BH mass remains the same. 
The isotropic case ejects more central gas outside the disk 
($\leq 1.3$ times the mass than in the bipolar case), 
because the wind kicked along the galaxy disk interact with more gas. 
Note that in our subgrid models the BHAR is the modified Bondi rate (Eq.~\ref{eq-Mdot-Bondi}), 
without any check on the radial velocity direction (\S\ref{sec-num-BH-Accr-Feed}). 

%%%%%%%%%%%%%%%%%%%%%%%%%%%%%%%%%%%%%%%%%%%%%%%%%%%%%%%%%%%%%%%%%%%%%%%% 

% FIGURE 9 
\begin{figure*} 
\centering 
\includegraphics[width = 1.0 \linewidth]{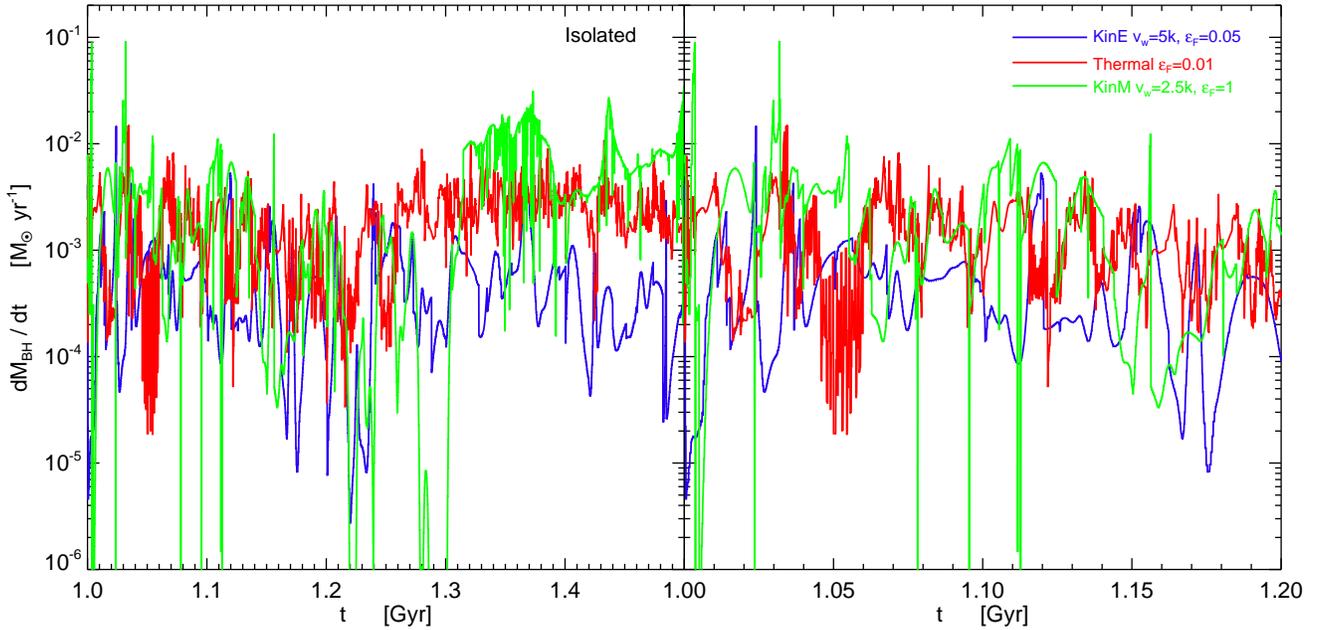} 
\caption{ 
Evolution with time of BH mass accretion rate in the fiducial isolated galaxy runs, 
showing zoomed-in time intervals: $(1 - 1.5)$ Gyr in the left panel, and $(1 - 1.2)$ Gyr in the right. 
The values at a finer time resolution (every timestep of simulation) are plotted, 
so that all variations are visible as spikes. 
The colours discriminate AGN feedback models as labelled: 
{\it th2}   ({\it red})   - thermal with $\epsilon_f = 0.01$, 
{\it kinE2} ({\it blue})  - kinetic EDW with $v_w =  5000$ km/s and $\epsilon_f = 0.05$, 
{\it kinM3} ({\it green}) - kinetic MDW with $v_w =  2500$ km/s and $\epsilon_f = 1$. 
} 
\label{fig-BHAccrRate-Zoom} 
\end{figure*} 

%%%%%%%%%%%%%%%%%%%%%%%%%%%%%%%%%%%%%%%%%%%%%%%%%%%%%%%%%%%%%%%%%%%%%%%% 

% period of intermittency of kinetic feedback. 

We find the BH mass accretion rate to be highly variable on a range of time scales. 
The top-left panel of Fig.~\ref{fig-BHAccr-StarForm-Rate} presents the BHAR over the 
whole evolution time $(0 - 2)$ Gyr, in our fiducial isolated galaxy simulations. 
Fig.~\ref{fig-BHAccrRate-Zoom} depicts the evolution through zoomed-in time intervals: 
$(1 - 1.5)$ Gyr in the left panel, and $(1 - 1.2)$ Gyr in the right. 
The BHAR values plotted in Fig.~\ref{fig-BHAccrRate-Zoom} are taken from every timestep 
of the simulations, 
hence are at a finer time resolution than in Fig.~\ref{fig-BHAccr-StarForm-Rate}. 
Therefore each and every variation is visible 
as a separate spike here in Fig.~\ref{fig-BHAccrRate-Zoom}. 
Overall, 
kinetic feedback causes larger amplitude of the fluctuations (variability factor between $10 - 10^4$) 
than thermal (up to $100$). 
At $t > 1$ Gyr with kinetic feedback, 
the accretion spikes of amplitude $\sim 100$ have a period of $\sim 0.05$ Gyr. 
Our results are consistent with other studies who find variability of the BHAR in simulations 
\citep[e.g.,][]{Gaspari12b, Gabor13}. 

We infer that different model parameters are needed to fit the $[M_{\rm BH} - \sigma_{\star}]$ 
observations for a BH evolving in an isolated galaxy versus that formed in a merger of two galaxies. 
Also varying modes of AGN feedback (thermal versus kinetic, additional details of wind prescription) 
require different parameter sets. 
BHs growing in cosmological environments undergo 
several major and minor mergers as well as quiescent evolution at different epochs. 
BHs also have alternate phases of feedback, quasar-mode versus radio-mode 
(which are usually realized as thermal or kinetic forms of feedback in simulations), 
at varying points of their lifetimes likely dependent on environment as well. 
The selection of a unique parameter set for cosmological simulations is hence not straight forward. 
We find that the best-fit $\epsilon_f$ is larger in a merger than in an isolated galaxy, 
considering either thermal or kinetic feedback. 
Furthermore, 
a larger $\epsilon_f$ is needed in kinetic over thermal within each of isolated or merger case. 

The obtained trend of larger efficiency for kinetic feedback is consistent with studies using 
the value $\epsilon_f = 0.15$ in the quasar-mode \citep{Booth09, Dubois13a}, 
and $\epsilon_f = 1$ in the radio-mode \citep{Dubois13a}, 
which were calibrated to reproduce the \citet{Magorrian98} relations at $z = 0$. 
Our best-fit parameter set [$v_w = 10000$ km/s, $\epsilon_f = 0.25$] of BH kinetic feedback EDW 
falls within the $[M_{\rm BH} - \sigma_{\star}]$ best-fit category 
for both isolated galaxy and merger. 
We plan to explore such parameters in full cosmological simulations in the future. 

Our results indicate that faster winds require higher $\epsilon_f$ 
to produce the same BH mass. 
This is because of the analytical dependence of the mass outflow rate on other parameters. 
E.g. for EDW of two different velocities having the same $\dot{M}_{\rm BH}$ and $\dot{M}_w$, 
equating the factor $( 2 \epsilon_f \epsilon_r c^2 / v_w^2 )$ in Eq.~(\ref{eq-MdotW-EDW}) 
gives: $\epsilon_{f1} / v_{w1}^2 = \epsilon_{f2} / v_{w2}^2$. 
Therefore, faster winds ($v_{w2} > v_{w1}$) need higher feedback efficiency 
($\epsilon_{f2} > \epsilon_{f1}$), to hold the previous equality. 
Our simulations explore EDW of: $v_{w1} = 5000$ and $v_{w2} = 10000$ km/s. 
In order to have the same BH mass, this analytical estimate predicts, 
$\epsilon_{f2} = \epsilon_{f1} (v_{w2} / v_{w1})^2 = 4 \epsilon_{f1}$. 
The best-fit parameters that we obtain for the isolated galaxy are: 
$\epsilon_{f1} = 0.05$ and $\epsilon_{f2} = 0.25$, 
whose ratio ($= 5$) is very close to the analytical prediction ($= 4$). 

The slope of the observational $[M_{\rm BH} - \sigma_{\star}]$ correlation 
has been revised over the last decade owing to sample selection effects and continuous newer data. 
Results depend on whether or not barred and/or pseudo-bulge galaxies are included in the analysis, 
and statistical regression methods used to fit the data. 
The observations by \citet{Tremaine02, Gultekin09} based on which we calibrate our model parameters 
have a $[M_{\rm BH} - \sigma_{\star}]$ logarithmic slope of $\sim 4$. 
In a relatively recent study \citet{McConnell13}, 
using new and revised kinematic data of a larger sample, 
presented significantly steeper $[M_{\rm BH} - \sigma_{\star}]$ relation of slope $\sim 5$. 
This steepening has occurred because of newest dynamical measurements of BH mass. 
Using observations of barred and non-barred galaxies, \citet{Graham11, Graham13} also found a steeper 
$[M_{\rm BH} - \sigma_{\star}]$ slope of around $5 - 5.5$ for elliptical and unbarred galaxies. 
\citet{Kormendy13} recently reported a slope of $4.4$ for $[M_{\rm BH} - \sigma_{\star}]$, 
almost midway between the values $4$ and $5$ obtained in other studies. 
Studying semi-analytical models, 
\citet{Shankar12} showed that the scatter in the updated local BH - bulge mass relation 
appears to be quite large when including late-type galaxies. 
\citet{Sadoun12} inferred observational evidence that the BH mass correlates 
with the velocity dispersion of globular cluster systems in their host galaxies; 
the relation having a flatter slope, a higher normalization, and less scatter. 

% The classical (all morphological type) relation for predicting black hole masses 
% has a slope of 5.13 \citep{Graham11}. 
% Alister also explained some of the observational criticalities in his latest reply. 

% (with no restriction on galaxy type) 
% in early-type galaxies (higher $\sigma_{\star}$) 
% mostly fall above the global $[M_{\rm BH} - \sigma_{\star}]$ relation, 
% and the newest measured $M_{\rm BH}$ in late-type galaxies (lower $\sigma_{\star}$) 
% mostly fall below the global scaling relation. 

% Here we compile an updated sample of 72 black holes and their host galaxies, 
% between BH mass and stellar velocity dispersion. 

%> Particle resolution and its effects are particularly important
%> when considering rapidly varying quantities, such as the SMBH
%> accretion rate, and making statements about duty cycles, etc.; how
%> much effect does particle noise have on the plots in Fig. 4? 

% a reason for which is an extra $6 x 10^7 Msun$ gas is converted to stars leaving less gas to be accreted. 
% at t < 0.7 Gyr more mass flows out of the disk at hi-res, 
% Attempting to find model parameters for an absolute fit at any resolution 
% Our goal in this work is to compare different AGN feedback modes to study their relative impact. 
% which is beyond the scope and do not constitute the primary goal of this paper. 

We perform one isolated galaxy simulation ({\it kinE2} case) 
with $10$ times higher resolution (run {\it hi-res}), to test resolution effects. 
We find that the {\it hi-res} results are qualitatively similar to that of {\it kinE2}, 
but not technically converged. 
The SFR depends on the resolved gas density, 
while BH growth depends on SFR through the local gas properties, 
because SF depletes gas, reducing that available for BH accretion. 
Therefore full convergence cannot be expected with the same parameter values of the subgrid models. 
The SFR is more in the {\it hi-res} case, since higher densities are resolved, 
and the mass of new stars formed is $\sim 1.5 \%$ larger, consequently 
the BH is $5$ times less-massive at $2$ Gyr, and less gas mass outflows after $0.7$ Gyr. 
The lack of convergence of the BH mass implies that different 
model parameters (e.g. $\epsilon_f, v_w$) are required to have the same 
$[M_{\rm BH} - \sigma_{\star}]$ fit at different resolution. 
Here a rigorous numerical convergence would first require tuning of the SF model parameters. 
E.g., \citet{Guedes11} found that their SF prescription requires 
higher density thresholds at increased resolutions. 
We find that the fluctuating nature of the 
rapidly varying quantities (BHAR and SFR in Fig.~\ref{fig-BHAccr-StarForm-Rate}) 
remain the same between {\it kinE2} and {\it hi-res} cases. 

Our results demonstrate that the current implementation of BH thermal feedback 
has essentially no effect within the framework of the standard multiphase star-formation model \citet{SH03}, 
where SF is based on a density threshold only. 
The thermal energy deposited to the gas particles which are multiphase (star-forming) 
is radiated away instantaneously, since they are dense. 
And they attain the effective equation-of-state temperature dictated by their density. 
This process might even induce SF by heating, 
since increasing the temperature of a multiphase particle makes the cold-phase more pressurized. 
This mimics the positive AGN feedback phenomenon of 
triggering starbursts by AGN outflows, when they compress clumpy gas clouds 
\citep[e.g.,][]{Natarajan98, Mellema02, Barai07, Gaibler12}. 

% overpressure and fragment 
% Our results imply that SF quenching happens by removal of gas, 
% and there is very little effect of feedback by gas heating on SFR. 

The non-effectiveness of the thermal feedback is aggravated in our models by the hole detection 
numerical algorithm (\S\ref{sec-num-HolePrevent}), first used in this study. 
The SFR in our simulations is directly proportional to the gas mass inside $2 h^{-1}$ kpc, 
implying that SF quenching happens by removal of gas. 
We extrapolate that all the studies \citep[e.g.,][]{DiMatteo05} 
using the BH numerical methodology of \citet{SDH05} are plagued by such artefacts: 
those incur a thermal feedback induced SF quenching by gas depletion 
via the creation of a hole around the BH location, and not actually by heating the gas. 
However we limit the expansion of the hole using our novel numerical technique, 
which restricts the gas depletion, 
therefore allows SF to occur unrestricted, bounding the impact of SF reduction. 

Note that \citet{Booth09} employed schemes to solve these problems, 
and the improved BH subgrid model has been used in their subsequent 
cosmological simulations \citep[e.g.,][]{Schaye10, vandeVoort11, Haas13}. 
\citet{Booth09} showed analytically that when $\alpha = 100$ is adopted, 
the BHAR of massive BHs is Eddington-limited (i.e. independent of density) down to very low gas density, 
and drop below Eddington only at even lower densities. 
They introduced a density-dependent $\alpha$ factor in the Bondi rate, 
which enables a BH to lower its accretion rate at higher gas densities, 
overall reducing the possibility of creating an artificial hole in the gas distribution. 
Furthermore \citet{Booth09} made thermal AGN feedback efficient by setting 
a minimum heating temperature of $10^8$ K, 
and allowing strongly heated gas (even if denser than SF threshold) 
to leave the effective SF equation-of-state. 

% The recent simulation result of \citet{Haas13} shows signature of 
% very strong AGN accretion / feedback such that the whole galaxy disk is destroyed 
% in the bottom-right panel of Fig. 2. 

Modifications are needed in the BH thermal feedback subgrid model in order to make it effective. 
One amendment is to not allow the thermally heated gas to form stars, 
to ensure the negative nature of AGN feedback. 
In subgrid models this amount to imposing an upper temperature threshold for SF, 
such that only gas at lower temperatures form stars \citep[e.g.,][]{Vogelsberger13}. 
An alternative method for BH thermal energy distribution 
is described in appendix A4 of \citet{Ragone-Figueroa13}. 
% which has been tested to be more effective than the default \citet{SDH05} model. 
The efficient implementation of thermal SNe feedback in galaxy simulations is a related issue; 
\citet{DallaVecchia12} prescribed a scheme where the gas is 
heated to a minimum heating temperature and shows strong feedback effects. 

The mass of newly formed stars together with the initial disk and bulge stars, 
plotted in the bottom row of Fig.~\ref{fig-XY-YZ-Isolated}, is indistinguishable between 
the three runs ({\it SF}, {\it th1}, {\it kinE1}). 
We checked that at $t = 1.73$ Gyr, 
the mass of new stars formed is the same for the no-BH and thermal feedback cases, 
and 0.8 times smaller in the kinetic feedback model. 
However this new star mass ($\sim 3.7 \times 10^{9} M_{\odot}$) is $12$ times smaller 
than the combined disk and bulge stellar components present in the initial galaxy models 
($4.7 \times 10^{10} M_{\odot}$, Table~\ref{Table-Galaxies}). 
Thus the total stellar mass is dominated by the old stars of disk and bulge, 
and the newly formed ones contribute just a small fraction. 

% cause the total stellar mass of the three models to be comparable in Fig.~\ref{fig-XY-YZ-Isolated}. 
% The time evolution of the total SFR displayed in Fig.~\ref{fig-BHAccr-StarForm-Rate}, bottom-left panel, 
% indicate that the star formation histories of the three models presented are quite similar. 
% Hence both the factors of similar SFR and 

The $Z_C$ profiles of the merged galaxy (Fig.~\ref{fig-Radial-Profiles-Merger}) reveal that 
the differences between the models are most prominent in the lower-mass and fiducial galaxies 
(of total masses $M_{\rm tot} / M_{\odot} = 1.40 \times 10^{11}$ and $1.12 \times 10^{12}$), 
where kinetic feedback results up to $(10 - 1000)$ times higher $Z_C$ than thermal, 
within $r \sim (20 - 100) h^{-1}$ kpc. 
Hence we infer that the CGM gas at such galactocentric distances can give 
the best $Z_C$ observational diagnostic to distinguish between various BH feedback models. 
Other studies also find that kinetic AGN feedback can uplift metals from galaxy to $100$s of kpc 
\citep[e.g.,][]{Gaspari11a}, 
making metallicity diagnostics an excellent way to differentiate feedback models. 

% The circumgalactic medium (CGM) gas at galactocentric distances $(20 - 100) h^{-1}$ kpc 
% are found to give the best carbon metallicity ($Z_C$) observational diagnostic 
% to distinguish between BH models, 
% where kinetic feedback results up to $(10 - 1000)$ times higher $Z_C$ than thermal. 

% A thermal bomb can not become a kinetic fast cold wind, 
% and a fast kinetic wind can not become a thermal bomb completely. 
% can accelerate cold clumps to high velocities on the way 
% Hence the discrimination between the two feedback modes can become ambiguous in certain regions. 

We note the possibility that parts of an outflow might effectively 
change type between kinetic and thermal. 
A kinetically-driven outflow, which is generally a fast wind, 
can shock and thermalize along its path in the central regions of the galaxy, 
and parts of it may turn to a slow thermal outflow. 
A thermally-driven wind, which is hot and isotropic, 
can evaporate cold clumps, generating fast kinetic wind components. 
However signatures of the original distinct feedback mode: 
isotropic thermal blast versus bipolar kinetic wind, remain different, 
which we try to diagnose in our study by the different galaxy properties. 

% The BH feedback-driven outflows retain their single phases, hence they are stable. 

Physical processes like fragmentation, instabilities and turbulence are important, 
occurring on small-scales within the galaxy ISM or in the outflows. 
However the resolution of our simulations is too low to model these phenomena. 
The gas particle mass is $1.8 \times 10^{5} M_{\odot}$ (Table~\ref{Table-Galaxies}) 
in our fiducial galaxy, and the gravitational softening length is $0.5 / h$ kpc. 
In order to study the effects of instabilities, sub-pc resolution is required. 
E.g. simulations by \citet{Barai12} (where the gas particle mass is $0.8 M_{\odot}$) 
and \citet{Gaspari13} found thermal-instability driven fragmentation. 
Thus our too coarse resolution renders it impossible to explore such small-scale processes. 
Intuitively, in the presence of 
hydrodynamic instabilities (such as Kelvin-Helmholtz and Rayleigh-Taylor) and/or thermal instability, 
the outflows would have lower velocities and higher mass outflow rates by entrainment of more gas. 
The morphological differences of the various feedback modes would then be somewhat reduced. 
However, over several Gyr, 
the BH mass is expected to remain the same via self-regulation.

\section{Summary and Conclusion} 
\label{sec-conclusion} 

We investigate different modes of AGN feedback in galaxy simulations. 
We examine two physical ways in which the feedback energy from a BH is coupled to the neighbouring gas: 
{\it thermal} - where the temperature is increased, and 
{\it kinetic} - where the velocity is boosted. 
We formulate kinetic feedback models   
with two free parameters: feedback efficiency and AGN wind velocity. 
The models are implemented in the TreePM - SPH code {\sc GADGET-3}: 
gas particles are stochastically selected from the neighbours 
and imparted an one-time $v_w$ boost. 
We incorporate a novel numerical algorithm 
to detect the existence of a hole in the gas distribution around the BH, and prevent its expansion. 
The code includes additional sub-resolution physics: 
metal-dependent radiative cooling and heating;   
star formation; stellar evolution and chemical enrichment.  % using a fixed stellar IMF. 

% $1.40 \times 10^{11} M_{\odot}$, % $8.93 \times 10^{12} M_{\odot}$. 
% and two more cases of mass $10$ times smaller and larger. 
% a rotationally supported gaseous and stellar disk, and a central stellar bulge. 
% Morphological examination reveals that 
% kinetic feedback EDW might promote the formation of star-forming gas spiral arms. 
% The gravitational softening lengths are $0.5 h^{-1}$ kpc for gas and stars, and $1 h^{-1}$ kpc for DM and BH. 

We perform simulations of isolated and merging disk galaxies, 
of total mass $1.12 \times 10^{12} M_{\odot}$ (similar as the Milky-Way) in our fiducial case. 
Each initial galaxy model contains $3 \times 10^5$ dark matter, 
$5 \times 10^4$ disk gas, $25 \times 10^3$ disk star, and $25 \times 10^3$ bulge star particles. 
New stars form during the simulation, 
and in our analysis of $\sigma_{\star}$ all the stars are counted. 
The collisionless BH particle has an initial dynamical mass $1.6$ times the DM particle mass 
and a seed mass of $10^5 M_{\odot}$. 
For the merger simulations, two equal-mass isolated galaxies 
are set on a parabolic collision course. 
We perform runs with the same non-AGN physics, and varying the AGN feedback models: 

\noindent - star-formation and chemical enrichment only (no BH), 

\noindent - thermal BH feedback, 

\noindent - kinetic BH feedback with energy-driven wind prescription, 

\noindent - kinetic BH feedback with momentum-driven wind.  

\noindent 
The results are summarized below. 

\begin{itemize} 

\item 
We compare the $M_{\rm BH}$ versus $\sigma_{\star}$ obtained in our simulations 
with observational data. 
All the stars lying within the stellar half-mass radius are tracked, 
and the median stellar velocity dispersion is estimated from hundred random line-of-sight directions. 
The parameters giving a best-fit in the isolated galaxy are: 
for BH thermal feedback $\epsilon_f = 0.01$; 
for BH kinetic EDW  $v_w = 5000$ km/s with $\epsilon_f = 0.05$, 
and $v_w = 10000$ km/s with $\epsilon_f = 0.25$; 
for BH kinetic MDW $v_w = 2500$ km/s with $\epsilon_f = 1$. 
We obtain the following best-fit parameters in the galaxy merger: 
for BH thermal $\epsilon_f = 0.05$; 
for BH kinetic EDW $v_w = 5000$ or $10000$ km/s with $\epsilon_f = 0.25$. 
For BH kinetic MDW feedback, none of the parameters we explored fit the observations; 
the BH mass is always too large. 
Our best-fit model parameters are dependent on simulation resolution, 
because SFR depends on the resolved gas density, affecting BH growth. 

\item 
In the isolated galaxy, there is no gas outflow in the star-formation only case. 
BH thermal feedback produces a late weak outflow, with $10^{-3}$ of all gas ejected after $1$ Gyr. 
BH kinetic feedback produces gas outbursts from $0.2$ Gyr, 
as bipolar jet-like outflows visible intermittently, separated by longer quiescent intervals, 
with $0.03 - 0.07$ of the gas mass outflown by $1$ Gyr. 
This results in a smaller stellar mass. 

% but continuous % , some of which slows down with time, reverses and inflows. 
% is the star-forming gas from the central 1 kpc, hence 

\item 
In the merging galaxy pair, collisions, shock heating and tidal interactions cause 
significant gas outflow from the beginning, even in the star-formation only case. 
BH models induce enhanced outflow: 
in thermal feedback up to $2$ times higher than star-formation only, 
and in kinetic feedback $5 - 8$ times more. 
The galaxies pass through a first and second pericenters at $0.4$ and $1.9$ Gyr, 
and coalesce between $(2 - 2.5)$ Gyr. 
The resulting merged galaxy has an extended, diffuse, spheroidal gaseous halo. 
Stellar and gas distribution are more spherically shaped in the star-formation only run, 
irregular to elliptically shaped with kinetic feedback, and intermediate with thermal. 

\item 
The BH mass growth occurs in a qualitatively similar manner for all the models: 
slow growth initially, exponential growth from $t \sim 0.5$ Gyr 
until $1$ Gyr in the isolated galaxy and until $1.5$ Gyr in the merger 
when its mass increases by a factor $10$ $-$ a few $100$ depending on the AGN model, 
and an almost steady-state afterward. 
The final $M_{\rm BH}$ is inversely proportional to $\epsilon_f$, and directly to $v_w$. 
The impact of varying $v_w$ is reduced in a merger than an isolated galaxy. 

% increasing the mass $(2 - 3)$ times up to $t \sim 0.5$ Gyr, subsequent 

\item 
The BH mass accretion rate exhibit heavy fluctuations, by a factor of up to 
$100$ in the isolated and $1000$ in the merger within $0.02$ Gyr. 
 
% an initial burst, then sudden decrease, 
% linear increase from $0.02$ Gyr until $0.7$ and $1$ Gyr in the isolated and merger cases. 
% The isolated galaxy BHAR remains at roughly steady state after $0.7$ Gyr until $3$ Gyr; 
% the maximum BHAR of the runs varying between $(0.005 - 0.1) M_{\odot}$/yr. 
% The merger BHAR display local peaks at different times; 
% one at $2$ Gyr during the second pericenter passage 
% marks a BHAR between $(0.1 - 2) M_{\odot}$/yr in the different models. 

\item 
The models display similar star formation rate in the isolated galaxy: 
at $t > 0.9$ Gyr kinetic feedback generates a $1.5$ times lower SFR than the 
star-formation only and thermal cases. 
In the merger, 
thermal feedback produces a $(1.5 - 10)$ times lower SFR than the star-formation only run at $t > 2$ Gyr. 
Kinetic feedback causes a greater suppression, starting from $t > 1$ Gyr, 
$5 - 100$ times lower SFR than the star-formation only case. 

We see that the SFR is directly proportional to the gas mass inside $2 h^{-1}$ kpc, 
implying that SF quenching happens by removal of gas, 
and there is very little effect of feedback by gas heating on SFR. 

% shows an initial burst and sudden decrease (same as BHAR). 

\item 
The temperature radial profiles of the merged galaxy display a local peak in the outer regions, 
which for the star-formation only and thermal feedback occurs at a radius $10$ times smaller 
than the kinetic feedback models. 
Gas density and carbon metallicity profiles demonstrate that 
kinetic feedback expels dense metal-rich gas out from central regions of galaxies, 
and enrich the lower-density CGM and IGM to $> 100$ kpc. 
Radial $Z_C$ profiles 
present most prominent differences between the models in galaxies of total masses 
$1.40 \times 10^{11}$ and $1.12 \times 10^{12} M_{\odot}$: 
kinetic feedback results up to $(10 - 1000)$ times higher $Z_C$ than thermal, 
within $r \sim (20 - 100) h^{-1}$ kpc; 
the CGM gas at such galactocentric distances can give the best $Z_C$ observational diagnostic 
to distinguish between BH feedback models. 

% is more effective than thermal, and % carry metals out of the galaxy 

\item 
The low to negligible impact of BH thermal feedback on gas properties 
reveal that our adopted methodology from \citet{SDH05} 
is ineffective within the framework of the multiphase star-formation model.  % \citep{SH03} 
It is enhanced by the fact that we limit the expansion of the hole 
using our novel numerical technique, which restricts gas depletion around the BH. 
Previous studies using the same BH thermal feedback model had SF quenching by gas depletion 
via the creation of an artificial hole, and not actually by heating the gas. 

% therefore allows SF to occur unrestricted, bounding the impact of SF reduction. 
% and it is difficult to perform a physical comparison with kinetic feedback. 

\end{itemize} 

We are performing ongoing work to improve the subgrid model of BH thermal feedback, 
by modifying the way in which feedback energy is distributed and changing the conditions of SF. 
In the future, we plan to perform cosmological hydrodynamic simulations 
including unified models of AGN feedback: quasar-mode and radio-mode 
(numerically implemented as thermal and kinetic feedback from BH), 
to study BH-galaxy coevolution and IGM properties.

\section*{Acknowledgements} 

We are most grateful to Volker Springel for allowing us to use the GADGET-3 code, 
and for his help with the initial conditions. 
PB thanks Pierluigi Monaco and Edoardo Tescari for useful discussions. 
Simulations for this paper were partly performed on the COSMOS 
Consortium supercomputer within the Dirac Facility jointly funded by 
STFC, the Large Facilities Capital Fund of BIS and the University of 
Cambridge, as well as the Darwin Supercomputer of the University of 
Cambridge High Performance Computing Service 
(http://www.hpc.cam.ac.uk/), provided by Dell Inc. using Strategic 
Research Infrastructure Funding from the Higher Education Funding 
Council for England. 
Remaining simulations were run at the CINECA Supercomputing Center, 
with CPU time assigned through an ISCRA Class-C proposal.  % , and through an INAF-CINECA grant. 
This work is supported by PRIN-MIUR 2009, PRIN-INAF 2009, INFN/PD51 grant. 
PB and MV are supported by the ERC Starting Grant ``cosmoIGM''. 
SB acknowledges support from the European Commission's Framework Programme 7 through 
the Marie Curie Initial Training Network ``CosmoComp''. % (PITN-GA-2009-238356). 

%%%%%%%%%%%%%%%%%%%%%%%%%%%%%%%%%%%%%%%%%%%%%%%%%%%%%%%%%%%%%%%%%%%%%%%%
%
%                   REFERENCES
% \clearpage 

%%%%%%%%%%%%%%%%%%%%%%%%%%%%%%%%%%%%%%%%%%%%%%%%%%%%%%%%%%%%%%%%%%%%%%%%%%%%%%%%%%%%%%%%%% 

\end{document}